\documentclass[superscriptaddress,floatfix,prl,twocolumn,amsmath,amssymb]{revtex4-2}

\usepackage[resetlabels,labeled]{multibib}
\usepackage{graphicx}
\usepackage{dcolumn}
\usepackage{bm}
\usepackage{amssymb}
\usepackage{amsmath}
\usepackage{mathtools}
\usepackage{color}
\usepackage{datetime}
\usepackage{footnote}
\usepackage{bbold}
\usepackage[normalem]{ulem}
\usepackage{dcolumn}
 \usepackage{ragged2e}
 \usepackage{xfrac}
 \usepackage{sidecap}
 \usepackage{setspace}
 \usepackage{multirow}
 \usepackage{xcolor}
 \usepackage{flushend}
\usepackage{hyperref}
\usepackage{cleveref}
\usepackage{microtype}
\usepackage{siunitx}
\usepackage{caption}
\usepackage{subcaption}
\usepackage{braket}

\DeclareMathOperator{\tr}{tr}

\DeclareMathOperator*{\argmax}{arg\,max}
\DeclareMathOperator*{\argmin}{arg\,min}

\begin{abstract}
Understanding and controlling engineered quantum systems is key to developing practical quantum technology. However, given the current technological limitations, such as fabrication imperfections and environmental noise, this is not always possible. To address these issues, a great deal of theoretical and numerical methods for quantum system identification and control have been developed. These methods range from traditional curve fittings, which are limited by the accuracy of the model that describes the system, to machine learning methods, which provide efficient control solutions but no control beyond the output of the model, nor insights into the underlying physical process. Here we experimentally demonstrate a ``graybox'' approach to construct a physical model of a quantum system and use it to design optimal control. We report superior performance over model fitting, while generating unitaries and Hamiltonians, which are quantities not available from the structure of standard supervised machine learning models. Our approach combines physics principles with high-accuracy machine learning and is effective with any problem where the required controlled quantities cannot be directly measured in experiments. This method naturally extends to time-dependent and open quantum systems, with applications in quantum noise spectroscopy and cancellation.

\end{abstract}
\begin{document}

\title{Experimental graybox quantum system identification and control}

\author{Akram Youssry}
\address{Quantum Photonics Laboratory and Centre for Quantum Computation and Communication Technology, RMIT University, Melbourne, VIC 3000, Australia}

\author{Yang Yang}
\address{Quantum Photonics Laboratory and Centre for Quantum Computation and Communication Technology, RMIT University, Melbourne, VIC 3000, Australia}

\author{Robert J. Chapman}
\address{Quantum Photonics Laboratory and Centre for Quantum Computation and Communication Technology, RMIT University, Melbourne, VIC 3000, Australia}
\address{ETH Zurich, Optical Nanomaterial Group, Institute for Quantum Electronics, Department of Physics, 8093 Zurich, Switzerland}

\author{Ben Haylock}
\address{Centre for Quantum Computation and Communication Technology (Australian Research Council), Centre for Quantum Dynamics, Griffith University, Brisbane, QLD 4111, Australia}
\address{Institute for Photonics and Quantum Sciences, SUPA,
Heriot-Watt University, Edinburgh EH14 4AS, United Kingdom}

\author{Francesco Lenzini}
\address{Centre for Quantum Computation and Communication Technology (Australian Research Council), Centre for Quantum Dynamics, Griffith University, Brisbane, QLD 4111, Australia}
\address{Institute of Physics, University of Muenster, 48149 Muenster, Germany}

\author{Mirko Lobino}
\address{Centre for Quantum Computation and Communication Technology (Australian Research Council), Centre for Quantum Dynamics, Griffith University, Brisbane, QLD 4111, Australia}
\address{Department of Industrial Engineering, University of Trento, via Sommarive 9, 38123 Povo, Trento, Italy}
\address{INFN-TIFPA, Via Sommarive 14, I-38123 Povo, Trento, Italy}

\author{Alberto Peruzzo}
\address{Quantum Photonics Laboratory and Centre for Quantum Computation and Communication Technology, RMIT University, Melbourne, VIC 3000, Australia}
\maketitle

Quantum technology promises to deliver exponentially faster computation, provably secure communications, and high-precision sensing \cite{leuchs2019quantum}.
However, during the fabrication and operation of a quantum device, there are many factors that can significantly impact its functionality, requiring characterization and control techniques to achieve high-level performance.
Generally, we are interested in uncovering the unknown relation between the control and the Hamiltonian governing the device, and then utilizing this information to drive the device toward a desired target. Typical targets include unitary gates, a specific Hamiltonian, or certain output probability distributions.

Approaches that directly aim to control the quantum device without first identifying it, includes dynamical decoupling and dynamically-corrected gates \cite{CPMG,DS1,OC2, DS5}, as well as direct gradient-based optimization, such as the commonly used GRAPE algorithm \cite{khaneja2005optimal} and its variants \cite{de_Fouquieres_2011, Ciaramella_2015, PhysRevA.99.052327, PhysRevA.95.042318, caneva2011chopped, Haas_2019, PhysRevA.97.042122, PhysRevA.99.042327}. These techniques only work when the dependence of the Hamiltonian on the control is known, because they are based on optimizing the fidelity to some target with respect to control. In situations where this dependence is unknown, the fidelity (and in general other cost functions) and/or its gradient, can be computed iteratively from experimental data. The control is optimized after each iteration and directly applied to the system for the next iteration. The physical system becomes part of a feedback architecture for designing the pulses without a need for a model. Examples of this approach are in \cite{PhysRevA.99.042327, Li_2017, Chen_2020, Yang_2020}, and are sometimes referred to as ``learning quantum control'' \cite{Dong_2021}. Evolutionary algorithms, such as Genetic Algorithm (GA) have been also proposed \cite{Judson_1992}, 
as optimization techniques with the advantage of being gradient-free, and are more likely to find global minima. These techniques can also be applied with a known model or directly from experimental measurements.
Reinforcement learning methods \cite{sivak2022model, baum2021experimental, niu2019universal, erdman2022driving}, are also model-free and have been recently explored for the purposes of removing the reliance on assumptions on the physical system, and have been successfully applied to controlling quantum systems. 

In many situations, it is important to reconstruct a mathematical model of the system from experimental data, a process referred to as ``system identification''. An identified model can be used to compare the behaviour of a fabricated device to its design, or to understand the underlying noise process affecting the system.
As the model can predict the behaviour of the system, it can be used for control as well.

The traditional approach to characterizing and controlling physical devices is based on theoretical models of the underlying processes governing the relationship between input and output signals. Example of fitting data to a physical model include \cite{matthews_manipulation_2009, Erdman_2022}. These ``whitebox'' (WB) models are based on parameter estimation via curve fitting and can be computationally expensive, inaccurate, or incomplete. For example, they do not consider unexpected input parameters or dynamics. Moreover, the models commonly used to describe the dynamics of an open quantum system (such as Lindblad's master equation), is valid only under strict assumptions and approximations of the noise (such as Markovianity) and the control (such as being ideal impulses). These assumptions do not hold for many quantum platforms, making the use of fixed a-priori WB models inaccurate.  

To increase accuracy and remove some of the limitations of the WB method, supervised machine learning techniques are proving useful for modeling and controlling complex physical systems. In particular, techniques such as neural networks are, in many cases, superior if not the only viable approach. However, this approach, referred to as ``blackbox'' (BB), does not provide any information about the underlying physics of the system. 
Nonetheless, it has been used in many applications such as quantifying Non-Markovianity of quantum systems \cite{fanchini2021estimating}, characterizing qubits and environments \cite{flurin2020using, papivc2022neural, wise2021using, palmieri2021multiclass}, quantum control \cite{ostaszewski2019approximation, khait2022optimal, zeng2020quantum}, quantum error correction \cite{bausch2020quantum, baireuther2018machine}, optimization of experimental quantum measurements \cite{lennon2019efficiently, 10.1038/s41467-018-06847-1}, and calibration of quantum devices \cite{10.1103/physrevapplied.15.044003}. 
A closely-related approach to system identification is the ``Hamiltonian Learning'' problem \cite{wiebe2014quantum, wang2017experimental}, where a fixed time-independent Hamiltonian (control is fixed) is inferred from quantum measurements. 

In situations where the quantities of interest, such as Hamiltonians, unitaries, and noise operators, cannot be directly accessed from the model, the use of a hybrid WB-BB or ``graybox" (GB) approach allows to both identify and control these quantities beyond the measurable dataset. Following the standard control engineering definition \cite{GB_book}, the aim of GB models is to merge an abstract mathematical structure, such as a neural network, with physical laws. While direct control methods, such as GRAPE, are sometimes referred to as grayboxes, because they are data-driven and may rely on prior knowledge, these methods do not identify a useful mathematical description of the system. The graybox architecture provides access to any physical quantity available from the physical part (WB) of the model.

GB has been proposed to model electrical drift in quantum photonic circuits \cite{youssry2020modeling}, 
as well as open quantum systems subject to classical \cite{youssry2020characterization} and quantum noise \cite{youssry2022Multi}, covering the case of a time-dependent Hamiltonian problem.
The GB model was also applied in the context of noise detection in the presence of a spectator qubit that acts as a sensor of the environment \cite{youssry2021noise}, 
and to geometric quantum gate synthesis \cite{perrier2020quantum}.

While GB has been used experimentally to characterize superconducting qubits \cite{10.1103/prxquantum.2.040355}, to date, no quantum device has been characterised using a GB model where the identified model was then used to design optimal quantum control.

Here, we experimentally demonstrate how to model and control a quantum device using the GB architecture when the Hamiltonian dependence on the control is unknown. 
We report high-fidelity preparation of arbitrary unitaries and output probability distributions of a reconfigurable three-mode integrated photonic device and uncover the Hamiltonian dependence on the control. Our GB approach outperforms the traditional model fitting methods and can successfully prepare unitary operations, which  are not accessible from the structure of a BB. Our results show a promising approach to enhance quantum control by understanding the physical processes, and open the way to improve the engineering of quantum devices.

\begin{figure*}
     \centering
     \includegraphics[width=0.8\textwidth]{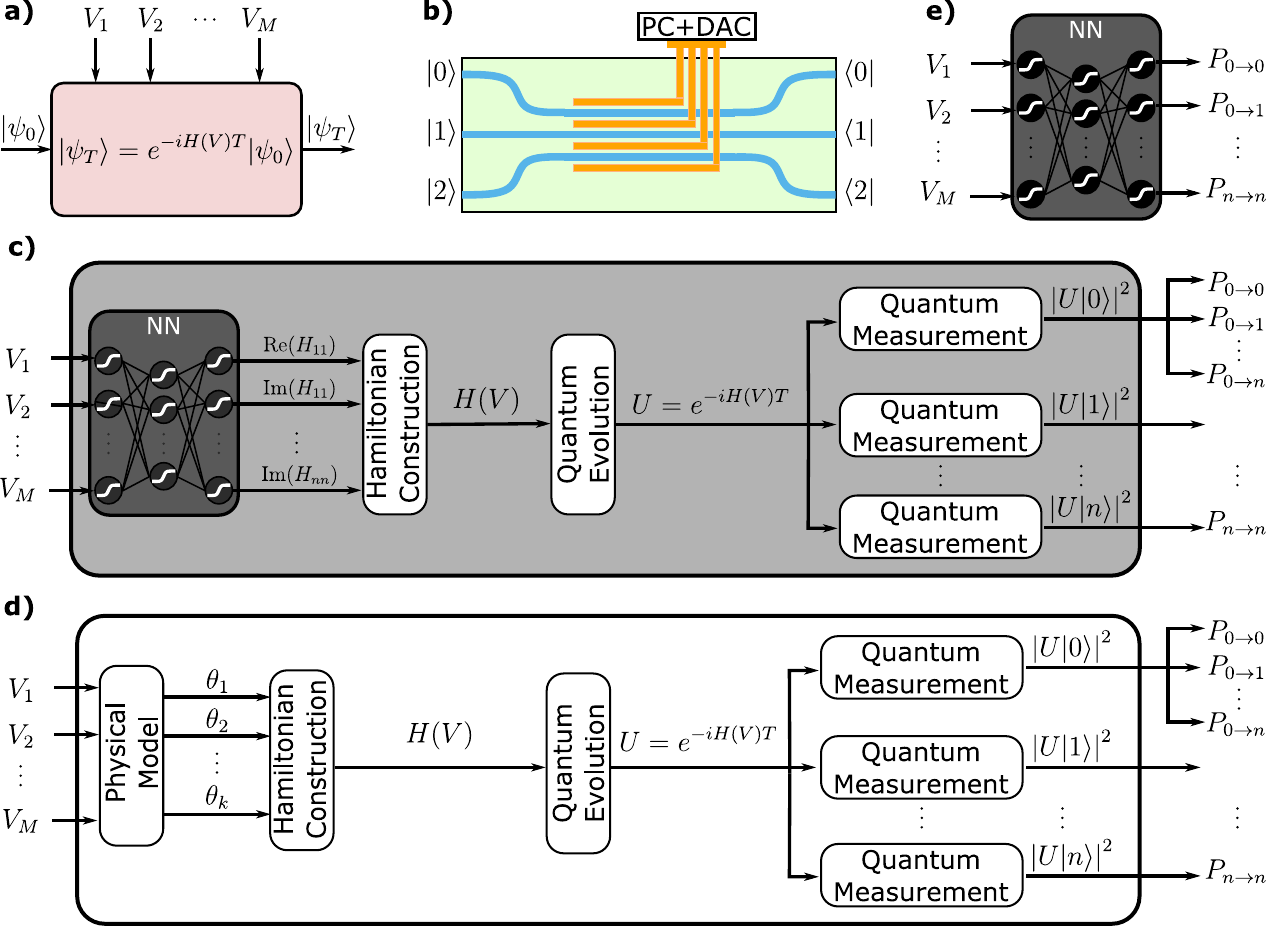}
    \caption{\textbf{Physical and machine learning models of the class of quantum devices considered in this paper}. These are described by a time-independent Hamiltonian in the absence of interaction with the environment. a) Representation of the class of quantum devices considered in this work. b) A schematic of an integrated photonic voltage-controlled reconfigurable waveguide array chip, implementing a noiseless time-independent Hamiltonian. Photons enter from the input port (on the left), undergo a voltage-controlled propagation along the chip, and are then measured at the output port of the chip (on the right). c) The structure of the proposed graybox model. The input to the model is the set of $M$ controls, while the outputs are the quantum measurements for the set of computational basis as initial states. $P_{a \to b}$ indicates the transition probability from input port $a$ to output port $b$. The  graybox is a combination of black and white boxes. The blackbox estimates the real and imaginary components of each matrix element of the Hamiltonian. The whitebox layers construct the Hamiltonian matrix and perform the quantum evolution and measurements. d) A fully whitebox architecture where a physical model is utilized. The first layer generates predefined Hamiltonian parameters that follow a known analytical dependence. The remaining layers perform the quantum evolution and measurements. e) A fully blackbox model where only a generic neural network is utilized with no physical model.}
    \label{fig:architecture}
\end{figure*}

\subsection{Modeling quantum devices}
We consider the class of quantum devices, shown in Fig.~\ref{fig:architecture}a, described by a time-independent Hamiltonian undergoing a closed-system evolution (i.e. in the absence of quantum noise). Photonic devices are examples of this class when the Hamiltonian is not modulated faster than the evolution time of a propagating photon. We focus on the case where the system is described by a finite $N$-dimensional Hilbert space (i.e. a qudit). The Hamiltonian governing the dynamics of the system can be represented in the most general form as an $N\times N$ complex Hermitian matrix that depends on a set of external controls. We encode the set of controls in a $M\times 1$-dimensional vector $\mathbf{V} = [V_1, V_2, \cdots V_{M}]^T$, where $V_k$ is the $k^{\text{th}}$ control. We assume that during the system evolution, the control vector is fixed. An example of such controls is the voltages applied to a reconfigurable photonic chip, schematically shown in Fig.~\ref{fig:architecture}b. The system starts in an initial state $\ket{\psi_0}$ and evolves to the state $\ket{\psi_T}$ at time $t=T$. The state is then measured on some basis to obtain a set of probability outcomes. A more general form of time-dependent evolution in the presence of unwanted interactions with the environment has been considered in our previous work \cite{youssry2020characterization}.

Our aim is to obtain a machine learning (ML) model that describes the behavior of the device given a set of controls $\mathbf{V}$ and use it for controlling the device. The input to the ML model is the $M$-dimensional control vector $\mathbf{V}$, while the output is the set of measured outcomes of the state after evolution. Generally, it is required to have informationally-complete measurements to fully characterize a quantum system. Here, we restrict the initial states as well as measurement basis to the set of computational basis states $\{\ket{0}, \ket{1}, \cdots \ket{N}\}$ in order to be compatible with our experimental setup. For each of these $N$ initial states, we have $N$ possible outcomes with an associated probability $P_{j\rightarrow k}$ corresponding to the $j^{th}$ input and $k^{th}$ output, giving a total of $N^2$ outputs. The approach, however, is independent of this choice, and any set of states can be used. In Supplementary Materials D, we discuss a more general measurement scheme. It is important to emphasize that the model input is the set of controls $\mathbf{V}$ applied to the system, and not the initial state $\ket{\psi_0}$. 

\subsection{Graybox architecture}

Our starting point is our theoretical proposal \cite{youssry2020modeling} for modelling and controlling quantum photonic circuits using a GB architecture. The work aimed at stabilizing the effect of electrical drift and preparing quantum gate sequences at the same time. In order to model such an effect, a GB was desgined to capture variations over a ``classical'' time scale (i.e. slower than the evolution time of a single photon). So, a recurrent neural network was used, particularly a Gated-Recurrent Unit (GRU) as the black part of the model. The inputs and outputs of the model are slowly time-varying waveforms. However, this resulted in optimal voltage pulses that did not belong to the class of pulses in the training set, which is not available in our experimental device, and may not in general be available.

Here, we focus only on modeling the unknown Hamiltonian-voltage dependence, and stabilize the drift in hardware using fast pulsing, a well known technique in integrated photonics. The pulses have fixed frequency and duty cycle, so the only controllable parameters are the amplitude on each electrode. This makes it possible to restrict the controller solution to the space of training pulses. Therefore, in what follows in this paper, we use standard feed-forward neural networks as opposed to recurrent neural networks as there is no need to model a sequence over time.

The GB structure we propose, shown in Fig.~\ref{fig:architecture}c, consists of a BB (in the form of a  neural network) followed by a WB part that processes the outputs of the BB into measurable physical quantities. The purpose of the blackbox is to map the controls $\mathbf{V}$ (the model inputs) to the Hamiltonian of the system. The output of the BB then represents the elements of the Hamiltonian matrix. A general $N\times N$ complex matrix has $2N^2$ degree of freedom ($N^2$ components with real and imaginary parts).

Thus, the output layer of the BB must consist of $2N^2$ neurons. The other BB layers can be designed arbitrarily, and are custom engineered to provide the best performance for a given dataset. 

The second layer is a Hamiltonian construction layer that arranges the outputs of the BB into an $N\times N$ matrix. A valid Hamiltonian has to be Hermitian (i.e. $H^{\dagger}=H$) and this is ensured by calculating the Hermitian part of the constructed matrix and discarding the anti-Hermitian part. This can be done simply by adding the matrix to its Hermitian conjugate. The output of this layer is a valid control-dependent Hamiltonian. We did not enforce any structure or constraint on the Hamiltonian, to enable the best fitting allowed by the rules of quantum mechanics for a given dataset.

The Hamiltonian is then followed by subsequent layers that transform the Hamiltonian matrix into the set of probability outcomes--which can be measured experimentally--utilizing the laws of quantum mechanics. Starting from a valid Hamiltonian, there is no need to further use a BB since the dynamical equations are known. This saves the algorithm from trying to learn the rules of quantum mechanics from experimental data, which would complicate the process and result in a less accurate model. We use WB layers for the remaining steps. In particular, there is a layer that calculates the evolution unitary by the matrix exponentiation of the Hamiltonian $U = e^{-i H T}$, which is the solution of the Schr\"{o}dinger's time-independent equation. The final part of the GB model is a concatenation of $N$-layers representing the quantum measurement operation for each of the $N$ input $\ket{j}$. In each layer we calculate $\ket{\psi_T} = U \ket{j}$, where $\ket{j}$ is the initial state of the system. After evolving the state to $\ket{\psi_T}$, the probabilities $P_{j\rightarrow k}$ are calculated by taking the absolute value squared of each entry of the state, that is applying the Born's rule for quantum expectation values. 

The added WB layers do not include any trainable parameter; they only exist in the BB part. Therefore, when we train the model on a dataset, the only updates occur in the BB, generating a set of outputs that can be interpreted as a Hamiltonian. 

An important aspect of the GB architecture is that it is independent of the physics of the system, i.e. it provides the most general form of a map between the control vector and the quantum measurements while keeping the most important physical quantities accessible via software, namely Hamiltonian, unitary and evolved state. This is the key aspect needed to perform quantum control as we are usually interested in implementing a quantum gate represented by the unitary or Hamiltonian and not by the measured evolved state. Having access to those quantities, even though the model is trained with quantum measurements, is the key feature of the GB architecture.

\begin{figure*} 
     \centering
    \includegraphics[width=0.9\textwidth]{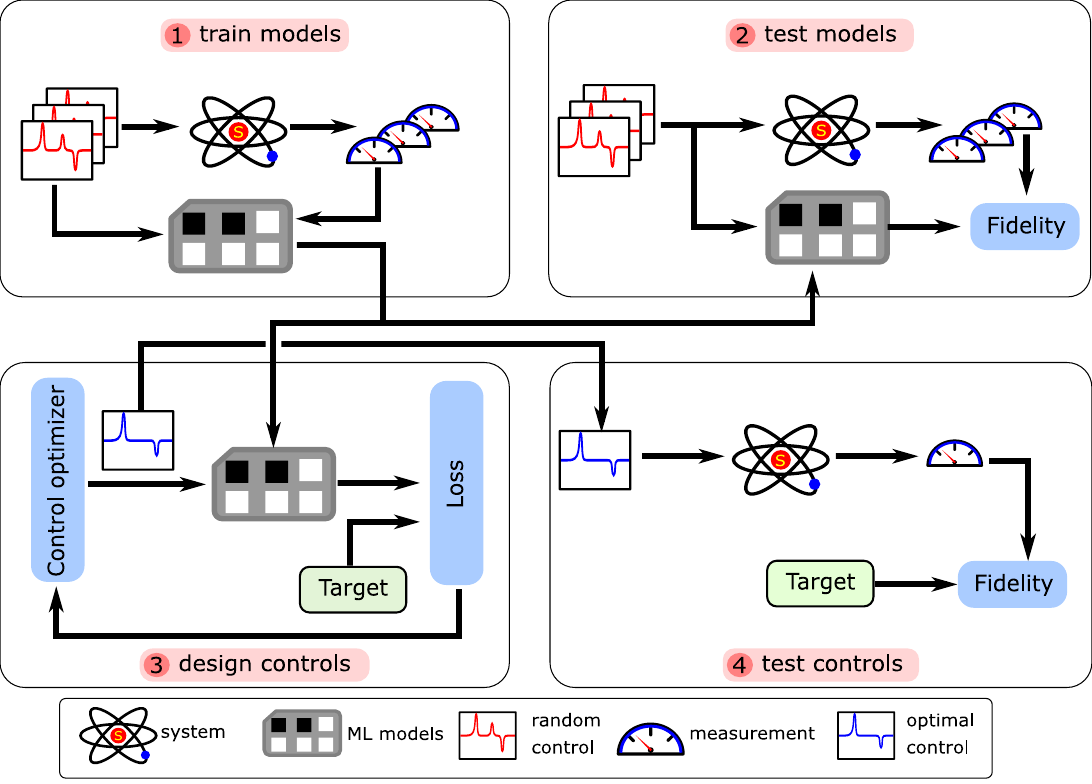}
    \caption{\textbf{Protocol schematic}. The first step is to construct an experimental dataset by applying controls to the system and measuring the corresponding outputs. The dataset is then used to train the machine learning models (1). Next, the trained models are tested (2) for generalization by comparing their output predictions against a different experimental testing dataset. After that, the trained models can be used to optimize controls (3) to achieve a certain target, which could be a Hamiltonian, a unitary gate, or an output probability distribution. Finally, the obtained controls are tested (4) experimentally and the controlled system output is compared against the desired target.}
    \label{protocol}
\end{figure*}

\subsection{Whitebox and blackbox architectures}
We benchmark the performance of our GB model against the fully WB and fully BB architectures. 
The WB approach (Fig.~\ref{fig:architecture}d) is equivalent to the standard model (curve) fitting and the details of the architecture depend on the physical system. The assumption is that all the relations between different dynamical variables are exactly known except for the parameters we are fitting. Generally, a WB consists of several layers. The first layer represents the mathematical relations between the controls and the Hamiltonian, with a set of unknown parameters. The input of this layer is the control vector, and the output is a mathematically valid Hamiltonian. The remaining layers are identical to those of the GB and represent quantum evolution and quantum measurements. For some systems, we might need more layers (e.g. to model the fan-in/out in photonic devices \cite{Peruzzo.2010.Science.10.1126/science.1193515j8}). The WB provides the same access to hidden quantities as the GB and even provides more physics as we know exactly the analytical relations between Hamiltonian and control. However, if we do not know these relations, or they do not match the physical reality, the WB will fail and will not be useful for further applications.

The BB architecture (Fig.~\ref{fig:architecture}e) is largely different from the WB model since the relation between the control vector and the quantum measurement is modeled by a neural network. While any structure can be used, in this paper we consider fully-connected neural networks with a softmax output layer of $N$ neurons. This enforces the outputs to form a probability distribution (i.e. positive numbers in $[0,1]$ whose sum is equal to $1$). This is consistent with what the model outputs represent, which are the probability amplitudes of the evolved quantum state. Since we are characterizing with $N$-initial basis states, we need $N$ of such layers with the initial state chosen accordingly. This will give a total of $N^2$ outputs. The structure of the other hidden layers can only be determined and optimized by examining the performance on an actual dataset. No other physical quantities can be accessed through this architecture, but it is very good in fitting a dataset since it gives the maximum freedom in terms of representation. If we are only interested in controlling the outputs, a BB would be an efficient solution. But, if the goal is to estimate physical quantities (like unitary gates), then a BB model is not useful at all. 

\subsection{Protocol for training, testing, and controlling}
Our protocol for training and testing models and controllers is schematically depicted in Fig.~\ref{protocol}.
It starts with preparing a dataset that will be used to train and test the machine learning BB. The dataset consists of examples. Each example is made of the $M$ control inputs $\mathbf{V}$ and the $N^2$ outputs of the model $P_{j\rightarrow k}$. So we start by generating random values for our control, let the system evolve, then perform the measurements and obtain the probabilities $P_{j\rightarrow k}$. We repeat this for the $N$ input states we consider to obtain all the outputs. The procedure is repeated for multiple examples. The number of examples of the dataset depends on the particular structure of the ML model, the noise level in the experiment, and the acceptable performance level. Generally, the larger dataset is, the better the ML algorithm will perform. 
In our previous work \cite{youssry2020modeling}, only computer simulated datasets were considered. In this paper, we create and test experimental datasets. This comes with many challenges including performing the experiment itself, the limited dataset size (to be feasible to collect), and the presence of noise not modeled by the quantum dynamics. In particular, initially, we found that statistical noise caused inconsistencies in the dataset, resulting in a poor performance of the method. As a result, we modified the dataset collection protocol, in particular the normalization of output power measurements as discussed in Supplementary Materials C. This new step improved the performance of the ML significantly.

After the dataset is collected, it is split into the training and testing subsets, and the ML model is trained. The purpose of training is to minimize a loss function that measures the distance between the predictions and actual outputs from the training dataset. In \cite{youssry2020modeling}, the loss function measured the similarity between waveforms, because the model was modeling drift, and the RMSprop algorithm \cite{rmsprop} was used for the training.  Here, we use the standard mean square error (MSE) as a loss function and use the ADAM algorithm \cite{Adam} for training.
Once the model is trained, its parameters are fixed and do not change in any of the remaining protocol stages. Next, the model is evaluated using the testing dataset, and its predictions are compared to the true corresponding values. The testing examples are not included in the optimization procedure, so they provide an unbiased evaluation of the performance of the model.  

Let's now consider the trained model for control purposes. In this case, the model acts as a replacement for the actual experimental setup and can be probed via software for any purpose. 
In \cite{youssry2020modeling}, the controller was designed to be a GRU, and therefore the optimal control was not restricted to any class of pulses. Here, we use a different controller that is designed to directly obtain the parameters of a fixed pulse shape (i.e. the voltage amplitudes). We consider two types of applications. The first is for obtaining the values of the control $\mathbf{V}$ to achieve a target output (i.e. probability amplitudes). In this case, we use the MSE as the control cost function, and the optimal control vector $\mathbf{V}^{*}$ can be expressed as 
\begin{align}
    \mathbf{V}^{*} = \argmin_{\mathbf{V}\in \mathcal{I}} (\hat{\mathbf{y}}_P(\mathbf{V}) - \mathbf{\mathbf{y}}_d)^T(\hat{\mathbf{y}}_P(\mathbf{V}) - \mathbf{y}_d)
\end{align}
where $\hat{\mathbf{y}}_P(\cdot)$ is the ML output predictions, $\mathbf{y}_d$ is the desired target, and $\mathcal{I}$ is the control domain, which reflects the maximum allowed range for the controls. To get accurate results, the ML model should be trained with dataset examples that lie in the same control domain as well. Note that the internal ML model parameters that define $\hat{\mathbf{y}}_P(\cdot)$ are not allowed to change during the optimization as they have been fixed after the training.

The other case is for achieving a target quantum gate (i.e. a target unitary). For this application, we use the gate fidelity as the control cost function for the controller defined as %
\begin{align}
    F(U,W) = \frac{\left|\tr{(U^{\dagger}W)}\right|^2}{N^2},
\end{align}
where $U$ and $W$ are two unitary matrices, and $N$ is their dimension. The gate fidelity lies in the range $[0,1]$, with $1$ representing the maximum overlap between the two gates. The optimal control voltages can then be represented as
\begin{align}
    \mathbf{V}^{*} = \argmax_{\mathbf{V}\in \mathcal{I}} F(\hat{y}_{U}(\mathbf{V}), U_d)
\end{align}
where $\hat{y}_U(\cdot)$ is the evolution unitary obtained from the ML model, and $U_d$ is the desired target gate. In this case, we can only use the GB and WB models, since a BB does not provide access to unitaries as discussed earlier. With this modular approach, the optimization algorithm of the controller can be chosen arbitrarily. While we choose a gradient-based method in this paper, other techniques such as genetic algorithms could also be used.

Finally, the optimal controls for a set of targets are applied experimentally, and the system is measured to construct the ``control" dataset. This dataset is then assessed and compared against the desired targets. Our main goal is to control a quantum system, and thus the assessment of any model should not just rely on its prediction capabilities, but also on how it performs in conjunction with a controller when tested experimentally.

\begin{figure*}
    \centering
    \includegraphics[width=2\columnwidth]{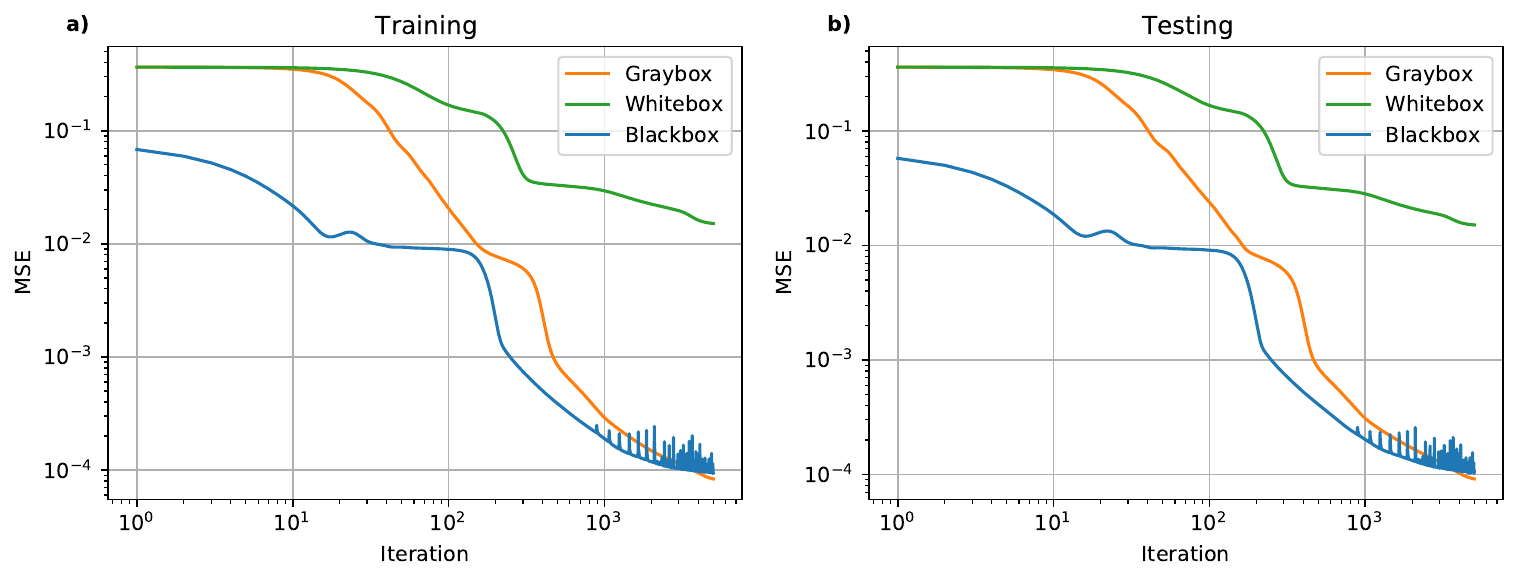}
    \caption{\textbf{Experimental performance of the machine learning models}. The whitebox model consists of fan-in, reconfigurable, and fan-out sections, each modelled as a real-valued tri-diagonal Hamiltonian in addition to a linear dependence on voltage for the reconfigurable section.  (a) Results of training the different models on the experimental dataset. The MSE is plotted versus iteration number (b) The results of evaluating the different models on the testing set.}
     \label{fig:training}
\end{figure*}

\subsection{Experimental results and discussion}
We tailor the design of the models described above around the device used for the experimental verification for our proposal, a voltage-controlled quantum photonic circuit of continuously coupled waveguides based on lithium niobate technology, 
schematically shown in Fig.~\ref{fig:architecture}b. The details about the chip's fabrication and its physical model are given in Supplementary Materials A and B. The chip has 3 waveguides, corresponding to a qutrit system, and is controlled by 4 electrodes and their respective voltages. In principle, there is no or negligible cross-talk in our device. This is guaranteed by the confinement of the electric field within the material due to the shielding effect from neighboring electrodes, as opposed to other technologies such as thermo-optic switching \cite{10011218}. Thus, the electrodes can be activated simultaneously, which is how we perform the experiments in this paper.
We implemented the ML models using the TensorFlow Python package \cite{tensorflow,keras}, applied the protocol for training the three models and then verified the performance of the controllers. The details of the implementations are also given in Supplementary Materials C. Moreover, we provide independent results of applying our method to a simulated synthetic dataset in Supplementary Materials D including a showcase for a 32-mode chip with 33 electrodes.

\begin{table}[b]
\begingroup
\begin{tabular}{|l|c|c|c|} 
\hline
    & \textbf{GB}   & \textbf{WB}   & \textbf{BB}\\ 
\hline
\multicolumn{1}{|l|}{\textbf{Model}}  & & &  \\ 
\quad average training MSE & $8.3\times10^{-5}$  & $1.5\times10^{-2}$ & $9.4\times10^{-5}$   \\ 
\quad average testing MSE & 
$9.1\times 10^{-5}$ & $1.5\times10^{-2}$ & $1.1\times10^{-4}$  \\ 
\textbf{Output Controller} & & & \\ 
\quad average MSE  & $2.6\times 10^{-3}$   & $1.7\times 10^{-2}$  &  $2.9\times 10^{-3}$ \\ 
\quad fidelity   &   &   &   \\ 
\qquad average &   $99.53\%$   &   $97.47\%$   & $99.48\%$ \\ 
\qquad with $>99\%$    &$87.3 \%$ & $34\%$  & $86.2\%$    \\ 
\textbf{Unitary Controller} & & & \\ 
\quad average MSE  & $3.1\times 10^{-3}$   & $1.9\times 10^{-2}$  &-- \\ 
\quad fidelity   &   &   &   \\ 
\qquad average & $99.48\%$ & $97.4\%$ &-- \\
\qquad with $>99\%$ & $71.3\%$ & $12\%$ &--\\
\hline
\end{tabular}
\endgroup
\caption{\textbf{Results summary}. Comparison of the performance of each model on the training and testing and control experimental datasets, as well as the experimental fidelities (average and instances greater than 99\%) for 1000 randomly prepared output distributions and unitary gates.}
\label{table1}
\end{table}

The results of the models training and testing as well as the control performance are reported in Table~\ref{table1}.
The MSE evaluated at each iteration for training and testing sets are shown in Fig.~\ref{fig:training}a and Fig.~\ref{fig:training}b. 
The plot of the learning curve in Fig.~\ref{fig:training}a shows the superior performance of the GB in terms of accuracy compared to the WB. This is due to the constraints imposed by the physical model that is used to construct the Hamiltonian in the case of the WB. As detailed in Supplementary Materials B,  the commonly-used device Hamiltonian is assumed to be tri-diagonal, real-valued, and linearly dependent on voltages. Our results show that these assumptions do not hold for a real device, and thus the degradation of the WB performance. 
On the other hand, the GB learns a general mathematically valid Hamiltonian and thus is able to better fit the experimental data. Initially, we designed the GB to enforce the Hamiltonian to be real-valued but otherwise arbitrary, and the fitting was not good. When we relaxed this assumption to allow a complex-valued Hamiltonian, the results were improved. Here we note that the Hamiltonian remains Hermitian to allow a unitary evolution, and thus it does not model losses. The normalization procedure (detailed in Supplementary Materials C) that we perform on the power measurements makes it unnecessary to model losses. The BB had similar performance to GB, with the main drawback of losing the physical picture.
In terms of the testing performance, Fig.~\ref{fig:training}b shows that the three models do not overfit, as the final MSE of the testing set is close to the final MSE of the training set. This means that the models do not memorize the examples of the training set. In other words, the model generalizes--there's no significant loss in prediction accuracy--confirming that the dataset, the model structure, and the training algorithm are well designed. Furthermore, the GB and BB clearly perform better than the WB. 

\begin{figure*}
    \centering
    \includegraphics[scale=0.7]{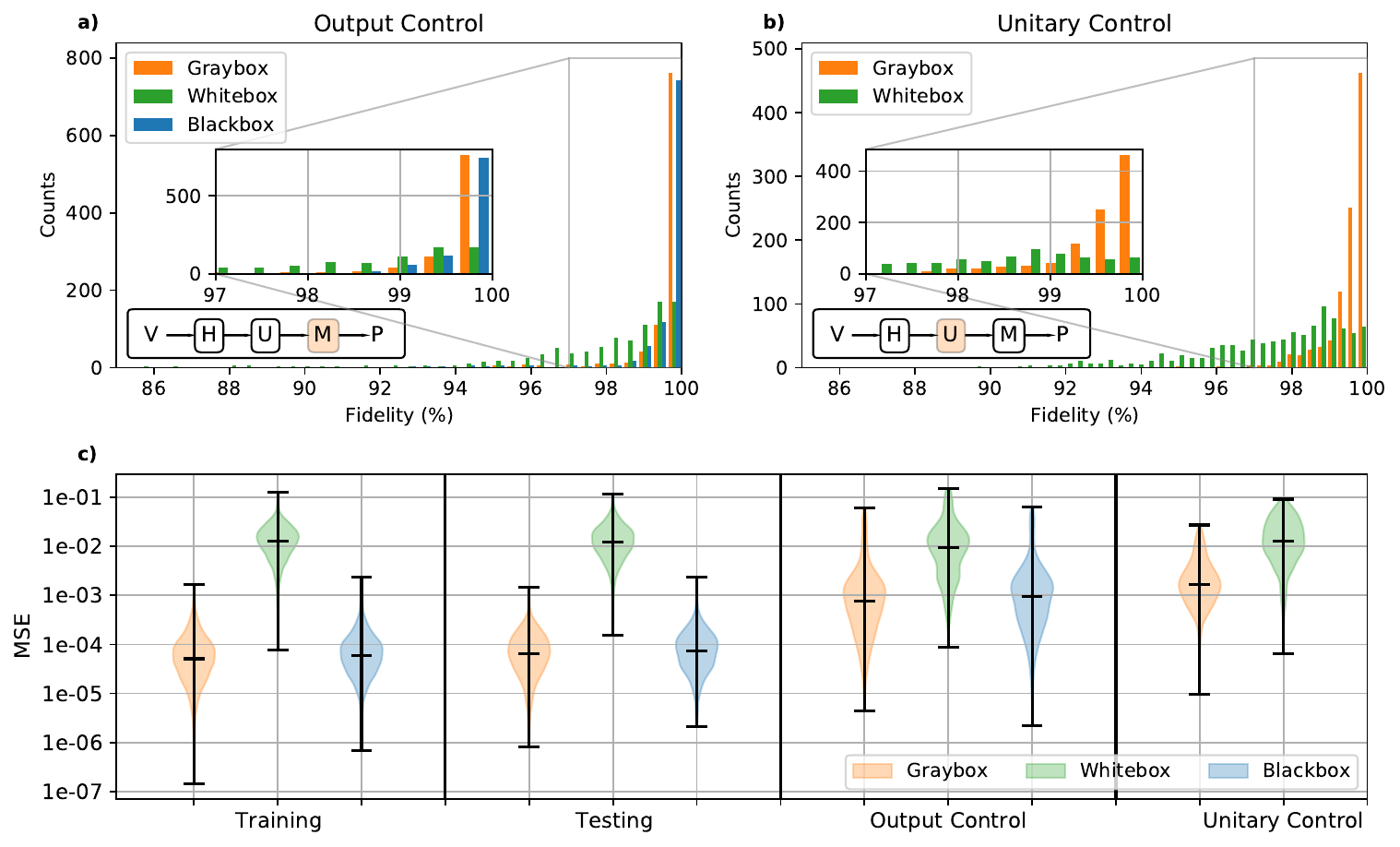}
    \caption{\textbf{Experimental quantum control performance}. The distribution of the fidelity between the experimentally measured output power distribution and the desired target distribution for the three models. The whitebox model utilizes a real-valued tri-diagonal Hamiltonian with linear dependence on voltages, in addition to fan-in and fan-out sections. The results are for (a) the output controller, and (b) the unitary controller. The reported values are the average over the three  distributions corresponding to each possible initial state. (c) Violin plot showing the statistics of the MSE obtained for the training, testing, and control datasets. The horizontal lines represent from bottom to top, the minimum, median, and maximum respectively. The plot also shows an estimated kernel density for the data.}
    \label{fig:results}
\end{figure*}

The performance of the models is limited by the size of the dataset. Because the experimental data always suffer from some level of noise, the minimum MSE obtainable without overfitting is limited as well. 
Usually, the acceptable level of MSE depends on the specific application. In this paper, our application is to control the chip to obtain target power distributions as well as target unitary operations. The ML models then act as a replacement/simulator of the actual setup. The experimental assessment of the optimal control will determine whether the model performance is accepted or need improvement. In general, the way to improve models is by constructing very large datasets which is the standard approach in most typical machine learning applications. For engineering applications, where we characterize and control a physical device, we are limited by how many measurements we can obtain. Thus, it becomes a tradeoff between the amount of time and resources needed to construct the dataset experimentally, and the accuracy of the trained models, which will also affect the performance of the controller.

The histogram of the controller fidelities between the desired target and the experimental measurements for 1000 randomly prepared output power distributions and 1000 randomly prepared unitary gates are shown in Fig.~\ref{fig:results}a and Fig.~\ref{fig:results}b respectively. The plots show that the WB is particularly skewed towards lower fidelities (minimum is $80.53\%$ compared to a minimum of approximately $91\%$ for GB and BB for the output controller). Similarly, for the gate controller, the minimum fidelity is $86.6 \%$ and $94.95 \%$ for WB and GB. In Fig.~\ref{fig:results}c, we summarize the statistics of the MSE between the ML model predictions and actual outputs for the training and testing datasets, in comparison with the MSE between the experimentally controlled measurements and the targets for each of the two controllers. 

The results of the controller for obtaining a target power distribution, show once again the superior performance of the GB and BB over the WB in terms of both MSE and average fidelity. The same controller/optimization algorithm and cost function are used for the three models. Thus, the lower performance comes from the lower accuracy of the WB model itself. When considering the controller for target unitary, it is only possible to use a WB or a GB as they are the only models that can give access to the overall unitary evolution matrix. A BB cannot be utilized in this case since it only encodes the dynamics in an abstract machine-suitable format, and does not provide any physical picture.

\begin{figure*}
    \centering
        \includegraphics[scale=0.6]{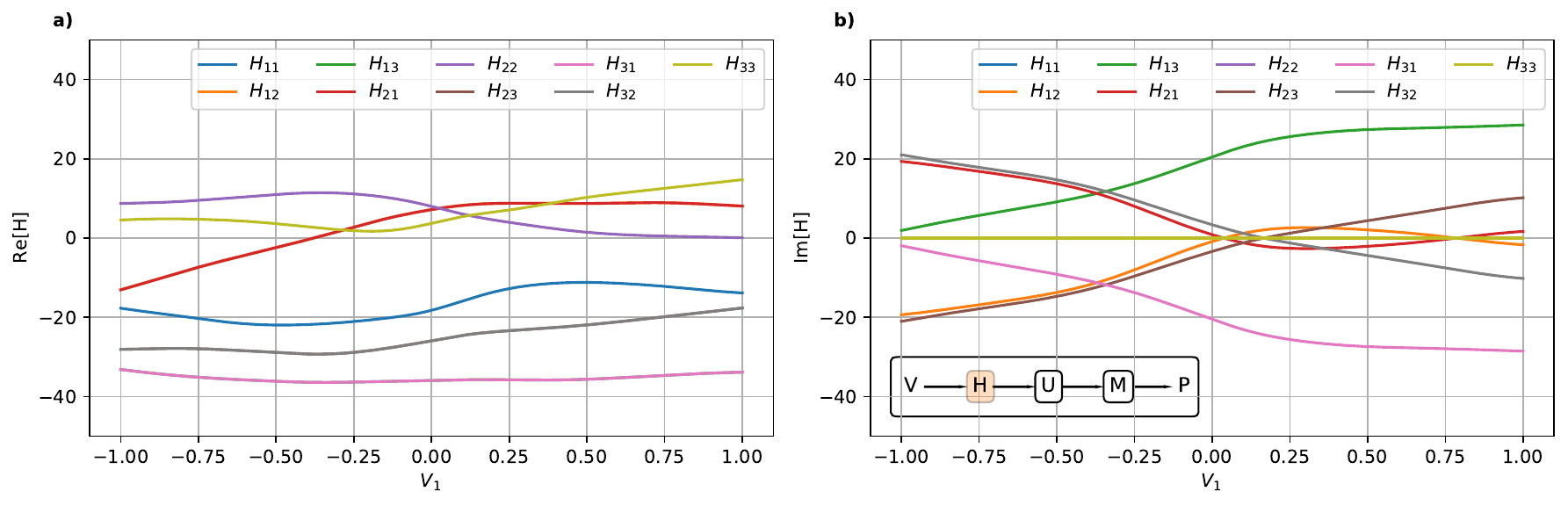}
    \caption{\textbf{Dependence of the Hamiltonian elements to a subset of input voltages, as predicted by the graybox model}. (a) Real and (b) imaginary parts of the Hamiltonian matrix elements as a function of voltage when all electrodes are grounded except the first electrode. It should be noted that the imaginary parts of $H_{11}$, $H_{22}$, and $H_{33}$ are by definition equal to zero. The non-linear dependence, the second off-diagonal elements, and the imaginary components indicate an effective Hamiltonian being estimated for a time-dependent system, attributed to non-homogeneity of the chip along the propagation direction.}
    \label{fig:Hamiltonian}
\end{figure*}

The performance assessment of ML-based algorithms on real rather than synthetic datasets is critical. Different noise sources could affect the data in unpredicted ways, which may also be difficult to simulate. This can affect the performance of the ML algorithm. We see that the final MSE of training and testing for the experimental dataset is two orders of magnitude less than that of the synthetic dataset (shown in the Supplementary Materials D). However, the control performance on the experimental dataset is accepted and thus we also accept the model prediction performance. In other situations, this might not be the case, and the ML design has to be modified to achieve higher performance. Therefore, using a design based on simulations such as \cite{youssry2020modeling} and applying it directly to an experimental dataset would not result in adequate performance, and so the whole workflow needs to be re-executed. 

Determining the required dataset size as well as NN architecture and complexity for higher-dimensional systems is generally difficult and has to be studied case by case. In Supplementary Materials D, we show promising results for a simulated 32-mode chip. And while the dataset and neural network sizes had to be increased, the overall protocol was still feasible to execute. 

Finally, the GB provides sufficient physical insights for most purposes, for example in Fig.~\ref{fig:Hamiltonian}a and \ref{fig:Hamiltonian}b, where the tunability of the Hamiltonian as a function of a single electrode voltage is explored. The figure shows the GB prediction of the different Hamiltonian elements as a function of the voltage applied to a single electrode. We can use these predictions more generally when more than one electrode is tuned (although it would be more difficult to plot in this situation) for unitary control. We can use the GB to predict the unitary given the set of voltages even though the dataset originally did not include this information, but rather the power distribution. Another advantage of using the GB compared to WB, is the incorporation of unmodelled effects such as cross-talk. While the effect is negligible in our technology, in other situations it can be difficult to have an exact/accurate WB model.

It is also important to realize that this predicted Hamiltonian, besides not being unique mathematically, represents physically an effective quantity, and so it will differ in structure (such as the existence of an imaginary part) from the ideal Hamiltonian that one may expect for a system. Depending on the purpose of the use of the GB we can control how much we make it ``blacker'' or ``whiter''. For control applications, the best architecture is to have this effective Hamiltonian. In another application such as modeling a device for the purpose of completely understanding the physics, the architecture of the GB might need to be modified to allow Hamiltonians that are closer to some expected structure. In Supplementary Materials E, we explore this idea in more detail. In particular, we explore the relaxation of the WB assumptions gradually until we reach the structure of the GB. The results show that the best architecture that fits the experimental data is a complex non-tridiagonal Hamiltonian with non-linear dependence on the control voltage. This suggests that the reconfigurable section of the chip has variations along the propagation direction, which could be the result of fabrication imperfections. In other words, the estimated Hamiltonian effectively represents a time-dependent system. Finally, it is worth mentioning that reaching this conclusion was based on interpreting the mathematical structure obtained from the GB. Thus, the GB approach helped us understand better the behaviour of our system.

\subsection{Conclusion and outlook}
We have shown how a GB model can be designed for a general quantum device, trained on experimental data, and verified by generating target unitary operations and output distributions with high fidelity. The performance was benchmarked against WB and BB models, showing the superior performance of our approach. 
Our approach is general and can be applied to any quantum system, it can be extended to time-dependent and open quantum systems with the needs of modifying the machine learning structure and dedicated dataset-taking process for specific hardware or quantum systems \cite{youssry2020characterization,youssry2021noise, youssry2022Multi}. There are many possible extensions to this work as well. One possibility is to design a GB for other physical models such as a Lindblad master equation for Markovian open quantum systems. One could also consider a hybrid approach between Hamiltonian learning using Genetic Algorithms (such as \cite{Schmidt.2009.Science, gentile2021learning}) and our numerical GB for the purpose of obtaining more detailed physical models. In terms of the ML-aspects of this application, a study about the scaling requirements for the NN structures of the GB in relation to the dimensionality of the system, would be interesting. However, it will be challenging because asymptotic analysis of ML algorithms is difficult or might be impossible. On the other hand, relying on numerical analysis might not be sufficient since the analysis will be restricted to a particular range of the scaling parameter, and cannot be generalized outside that range. Another aspect related to any ML algorithm is the requirements of the training dataset size. For complex devices, it can be challenging to collect a large-sized dataset. However, some emerging techniques can facilitate this process including incremental learning \cite{losing2018incremental}, transfer learning \cite{weiss2016survey}, and adaptive online learning \cite{hoi2021online}. While these techniques are constantly developing in classical machine learning literature, there is still a gap in porting such methods to physics-based applications and especially quantum applications.

\vspace{2mm}
\noindent\textbf{Data Availability}\\
The data generated in this study is available upon reasonable request from the corresponding author. 

\vspace{2mm}
\noindent\textbf{Code Availability}\\
The codes developed to generate this study is available upon reasonable request from the corresponding author. 
%
\vspace{2mm}
\noindent\textbf{ACKNOWLEDGEMENTS}\\
AP acknowledges an RMIT University Vice-Chancellor's Senior Research Fellowship and a Google Faculty Research Award. ML was supported by the Australian Research Council (ARC) Future Fellowship (FT180100055). BH was supported by the Griffith University Postdoctoral Fellowship. This work was supported by the Australian Government through the Australian Research Council under the Centre of Excellence scheme (No: CE170100012), and the Griffith University Research Infrastructure Program. This work was performed in part at the Queensland node of the Australian National Fabrication Facility, a company established under the National Collaborative Research Infrastructure Strategy to provide nano- and micro-fabrication facilities for Australia's researchers. This research was also undertaken with the assistance of resources from the National Computational Infrastructure (NCI Australia), an NCRIS enabled capability supported by the Australian Government.

\vspace{2mm}
\noindent\textbf{AUTHOR CONTRIBUTIONS}\\
A.Y. and A.P. conceived the experiment. B.H., F.L. and M.L.  fabricated the integrated photonic device. A.Y, Y.Y, R.J.C. and A.P. carried out the experiments and performed the data analysis. All the authors discussed the results and contributed to the writing of the paper.   

\vspace{2mm}
\noindent\textbf{COMPETING INTERESTS}\\
The Authors declare no Competing Financial or Non-Financial Interests.


\begin{thebibliography}{58}%
\makeatletter
\providecommand \@ifxundefined [1]{%
 \@ifx{#1\undefined}
}%
\providecommand \@ifnum [1]{%
 \ifnum #1\expandafter \@firstoftwo
 \else \expandafter \@secondoftwo
 \fi
}%
\providecommand \@ifx [1]{%
 \ifx #1\expandafter \@firstoftwo
 \else \expandafter \@secondoftwo
 \fi
}%
\providecommand \natexlab [1]{#1}%
\providecommand \enquote  [1]{``#1''}%
\providecommand \bibnamefont  [1]{#1}%
\providecommand \bibfnamefont [1]{#1}%
\providecommand \citenamefont [1]{#1}%
\providecommand \href@noop [0]{\@secondoftwo}%
\providecommand \href [0]{\begingroup \@sanitize@url \@href}%
\providecommand \@href[1]{\@@startlink{#1}\@@href}%
\providecommand \@@href[1]{\endgroup#1\@@endlink}%
\providecommand \@sanitize@url [0]{\catcode `\\12\catcode `\$12\catcode
  `\&12\catcode `\#12\catcode `\^12\catcode `\_12\catcode `\%12\relax}%
\providecommand \@@startlink[1]{}%
\providecommand \@@endlink[0]{}%
\providecommand \url  [0]{\begingroup\@sanitize@url \@url }%
\providecommand \@url [1]{\endgroup\@href {#1}{\urlprefix }}%
\providecommand \urlprefix  [0]{URL }%
\providecommand \Eprint [0]{\href }%
\providecommand \doibase [0]{https://doi.org/}%
\providecommand \selectlanguage [0]{\@gobble}%
\providecommand \bibinfo  [0]{\@secondoftwo}%
\providecommand \bibfield  [0]{\@secondoftwo}%
\providecommand \translation [1]{[#1]}%
\providecommand \BibitemOpen [0]{}%
\providecommand \bibitemStop [0]{}%
\providecommand \bibitemNoStop [0]{.\EOS\space}%
\providecommand \EOS [0]{\spacefactor3000\relax}%
\providecommand \BibitemShut  [1]{\csname bibitem#1\endcsname}%
\let\auto@bib@innerbib\@empty
\bibitem [{\citenamefont {Leuchs}\ and\ \citenamefont
  {Bruss}(2019)}]{leuchs2019quantum}%
  \BibitemOpen
  \bibfield  {author} {\bibinfo {author} {\bibfnamefont {G.}~\bibnamefont
  {Leuchs}}\ and\ \bibinfo {author} {\bibfnamefont {D.}~\bibnamefont {Bruss}},\
  }\href@noop {} {\emph {\bibinfo {title} {Quantum information: from
  foundations to quantum technology applications}}}\ (\bibinfo  {publisher}
  {John Wiley \& Sons},\ \bibinfo {year} {2019})\BibitemShut {NoStop}%
\bibitem [{\citenamefont {Carr}\ and\ \citenamefont {Purcell}(1954)}]{CPMG}%
  \BibitemOpen
  \bibfield  {author} {\bibinfo {author} {\bibfnamefont {H.~Y.}\ \bibnamefont
  {Carr}}\ and\ \bibinfo {author} {\bibfnamefont {E.~M.}\ \bibnamefont
  {Purcell}},\ }\bibfield  {title} {\bibinfo {title} {Effects of diffusion on
  free precession in nuclear magnetic resonance experiments},\ }\href
  {https://doi.org/10.1103/PhysRev.94.630} {\bibfield  {journal} {\bibinfo
  {journal} {Physical Review}\ }\textbf {\bibinfo {volume} {94}},\ \bibinfo
  {pages} {630} (\bibinfo {year} {1954})}\BibitemShut {NoStop}%
\bibitem [{\citenamefont {Viola}\ \emph {et~al.}(1999)\citenamefont {Viola},
  \citenamefont {Knill},\ and\ \citenamefont {Lloyd}}]{DS1}%
  \BibitemOpen
  \bibfield  {author} {\bibinfo {author} {\bibfnamefont {L.}~\bibnamefont
  {Viola}}, \bibinfo {author} {\bibfnamefont {E.}~\bibnamefont {Knill}},\ and\
  \bibinfo {author} {\bibfnamefont {S.}~\bibnamefont {Lloyd}},\ }\bibfield
  {title} {\bibinfo {title} {Dynamical decoupling of open quantum systems},\
  }\href {https://doi.org/10.1103/PhysRevLett.82.2417} {\bibfield  {journal}
  {\bibinfo  {journal} {Phys. Rev. Lett.}\ }\textbf {\bibinfo {volume} {82}},\
  \bibinfo {pages} {2417} (\bibinfo {year} {1999})}\BibitemShut {NoStop}%
\bibitem [{\citenamefont {Biercuk}\ \emph {et~al.}(2011)\citenamefont
  {Biercuk}, \citenamefont {Doherty},\ and\ \citenamefont {Uys}}]{OC2}%
  \BibitemOpen
  \bibfield  {author} {\bibinfo {author} {\bibfnamefont {M.~J.}\ \bibnamefont
  {Biercuk}}, \bibinfo {author} {\bibfnamefont {A.~C.}\ \bibnamefont
  {Doherty}},\ and\ \bibinfo {author} {\bibfnamefont {H.}~\bibnamefont {Uys}},\
  }\bibfield  {title} {\bibinfo {title} {Dynamical decoupling sequence
  construction as a filter-design problem},\ }\href
  {https://doi.org/10.1088/0953-4075/44/15/154002} {\bibfield  {journal}
  {\bibinfo  {journal} {Journal of Physics B: Atomic, Molecular and Optical
  Physics}\ }\textbf {\bibinfo {volume} {44}},\ \bibinfo {pages} {154002}
  (\bibinfo {year} {2011})}\BibitemShut {NoStop}%
\bibitem [{\citenamefont {Khodjasteh}\ and\ \citenamefont {Viola}(2009)}]{DS5}%
  \BibitemOpen
  \bibfield  {author} {\bibinfo {author} {\bibfnamefont {K.}~\bibnamefont
  {Khodjasteh}}\ and\ \bibinfo {author} {\bibfnamefont {L.}~\bibnamefont
  {Viola}},\ }\bibfield  {title} {\bibinfo {title} {Dynamically error-corrected
  gates for universal quantum computation},\ }\href
  {https://doi.org/10.1103/PhysRevLett.102.080501} {\bibfield  {journal}
  {\bibinfo  {journal} {Phys. Rev. Lett.}\ }\textbf {\bibinfo {volume} {102}},\
  \bibinfo {pages} {080501} (\bibinfo {year} {2009})}\BibitemShut {NoStop}%
\bibitem [{\citenamefont {Khaneja}\ \emph {et~al.}(2005)\citenamefont
  {Khaneja}, \citenamefont {Reiss}, \citenamefont {Kehlet}, \citenamefont
  {Schulte-Herbr{\"u}ggen},\ and\ \citenamefont {Glaser}}]{khaneja2005optimal}%
  \BibitemOpen
  \bibfield  {author} {\bibinfo {author} {\bibfnamefont {N.}~\bibnamefont
  {Khaneja}}, \bibinfo {author} {\bibfnamefont {T.}~\bibnamefont {Reiss}},
  \bibinfo {author} {\bibfnamefont {C.}~\bibnamefont {Kehlet}}, \bibinfo
  {author} {\bibfnamefont {T.}~\bibnamefont {Schulte-Herbr{\"u}ggen}},\ and\
  \bibinfo {author} {\bibfnamefont {S.~J.}\ \bibnamefont {Glaser}},\ }\bibfield
   {title} {\bibinfo {title}
  {\href{https://doi.org/10.1016/j.jmr.2004.11.004}{Optimal control of coupled
  spin dynamics: design of NMR pulse sequences by gradient ascent
  algorithms}},\ }\href {https://doi.org/10.1016/j.jmr.2004.11.004} {\bibfield
  {journal} {\bibinfo  {journal} {J. Magn. Reson.}\ }\textbf {\bibinfo {volume}
  {172}},\ \bibinfo {pages} {296} (\bibinfo {year} {2005})}\BibitemShut
  {NoStop}%
\bibitem [{\citenamefont {de~Fouquieres}\ \emph {et~al.}(2011)\citenamefont
  {de~Fouquieres}, \citenamefont {Schirmer}, \citenamefont {Glaser},\ and\
  \citenamefont {Kuprov}}]{de_Fouquieres_2011}%
  \BibitemOpen
  \bibfield  {author} {\bibinfo {author} {\bibfnamefont {P.}~\bibnamefont
  {de~Fouquieres}}, \bibinfo {author} {\bibfnamefont {S.}~\bibnamefont
  {Schirmer}}, \bibinfo {author} {\bibfnamefont {S.}~\bibnamefont {Glaser}},\
  and\ \bibinfo {author} {\bibfnamefont {I.}~\bibnamefont {Kuprov}},\
  }\bibfield  {title} {\bibinfo {title}
  {\href{https://doi.org/10.1016/j.jmr.2011.07.023}{Second order gradient
  ascent pulse engineering}},\ }\href
  {https://doi.org/10.1016/j.jmr.2011.07.023} {\bibfield  {journal} {\bibinfo
  {journal} {J. Magn. Reson.}\ }\textbf {\bibinfo {volume} {212}},\ \bibinfo
  {pages} {412} (\bibinfo {year} {2011})}\BibitemShut {NoStop}%
\bibitem [{\citenamefont {Ciaramella}\ \emph {et~al.}(2015)\citenamefont
  {Ciaramella}, \citenamefont {Borz{\`{\i}}}, \citenamefont {Dirr},\ and\
  \citenamefont {Wachsmuth}}]{Ciaramella_2015}%
  \BibitemOpen
  \bibfield  {author} {\bibinfo {author} {\bibfnamefont {G.}~\bibnamefont
  {Ciaramella}}, \bibinfo {author} {\bibfnamefont {A.}~\bibnamefont
  {Borz{\`{\i}}}}, \bibinfo {author} {\bibfnamefont {G.}~\bibnamefont {Dirr}},\
  and\ \bibinfo {author} {\bibfnamefont {D.}~\bibnamefont {Wachsmuth}},\
  }\bibfield  {title} {\bibinfo {title}
  {\href{https://doi.org/10.1137/140966988}{Newton methods for the optimal
  control of closed quantum spin systems}},\ }\href
  {https://doi.org/10.1137/140966988} {\bibfield  {journal} {\bibinfo
  {journal} {SICOMP}\ }\textbf {\bibinfo {volume} {37}},\ \bibinfo {pages}
  {A319} (\bibinfo {year} {2015})}\BibitemShut {NoStop}%
\bibitem [{\citenamefont {Abdelhafez}\ \emph {et~al.}(2019)\citenamefont
  {Abdelhafez}, \citenamefont {Schuster},\ and\ \citenamefont
  {Koch}}]{PhysRevA.99.052327}%
  \BibitemOpen
  \bibfield  {author} {\bibinfo {author} {\bibfnamefont {M.}~\bibnamefont
  {Abdelhafez}}, \bibinfo {author} {\bibfnamefont {D.~I.}\ \bibnamefont
  {Schuster}},\ and\ \bibinfo {author} {\bibfnamefont {J.}~\bibnamefont
  {Koch}},\ }\bibfield  {title} {\bibinfo {title}
  {\href{https://doi.org/10.1103/PhysRevA.99.052327}{Gradient-based optimal
  control of open quantum systems using quantum trajectories and automatic
  differentiation}},\ }\href {https://doi.org/10.1103/PhysRevA.99.052327}
  {\bibfield  {journal} {\bibinfo  {journal} {Phys. Rev. A}\ }\textbf {\bibinfo
  {volume} {99}},\ \bibinfo {pages} {052327} (\bibinfo {year}
  {2019})}\BibitemShut {NoStop}%
\bibitem [{\citenamefont {Leung}\ \emph {et~al.}(2017)\citenamefont {Leung},
  \citenamefont {Abdelhafez}, \citenamefont {Koch},\ and\ \citenamefont
  {Schuster}}]{PhysRevA.95.042318}%
  \BibitemOpen
  \bibfield  {author} {\bibinfo {author} {\bibfnamefont {N.}~\bibnamefont
  {Leung}}, \bibinfo {author} {\bibfnamefont {M.}~\bibnamefont {Abdelhafez}},
  \bibinfo {author} {\bibfnamefont {J.}~\bibnamefont {Koch}},\ and\ \bibinfo
  {author} {\bibfnamefont {D.}~\bibnamefont {Schuster}},\ }\bibfield  {title}
  {\bibinfo {title} {\href{https://doi.org/10.1103/PhysRevA.95.042318}{Speedup
  for quantum optimal control from automatic differentiation based on graphics
  processing units}},\ }\href {https://doi.org/10.1103/PhysRevA.95.042318}
  {\bibfield  {journal} {\bibinfo  {journal} {Phys. Rev. A}\ }\textbf {\bibinfo
  {volume} {95}},\ \bibinfo {pages} {042318} (\bibinfo {year}
  {2017})}\BibitemShut {NoStop}%
\bibitem [{\citenamefont {Caneva}\ \emph {et~al.}(2011)\citenamefont {Caneva},
  \citenamefont {Calarco},\ and\ \citenamefont
  {Montangero}}]{caneva2011chopped}%
  \BibitemOpen
  \bibfield  {author} {\bibinfo {author} {\bibfnamefont {T.}~\bibnamefont
  {Caneva}}, \bibinfo {author} {\bibfnamefont {T.}~\bibnamefont {Calarco}},\
  and\ \bibinfo {author} {\bibfnamefont {S.}~\bibnamefont {Montangero}},\
  }\bibfield  {title} {\bibinfo {title}
  {\href{https://doi.org/10.1103/physreva.84.022326}{Chopped random-basis
  quantum optimization}},\ }\href {https://doi.org/10.1103/physreva.84.022326}
  {\bibfield  {journal} {\bibinfo  {journal} {Phys. Rev. A}\ }\textbf {\bibinfo
  {volume} {84}},\ \bibinfo {pages} {022326} (\bibinfo {year}
  {2011})}\BibitemShut {NoStop}%
\bibitem [{\citenamefont {Haas}\ \emph {et~al.}(2019)\citenamefont {Haas},
  \citenamefont {Puzzuoli}, \citenamefont {Zhang},\ and\ \citenamefont
  {Cory}}]{Haas_2019}%
  \BibitemOpen
  \bibfield  {author} {\bibinfo {author} {\bibfnamefont {H.}~\bibnamefont
  {Haas}}, \bibinfo {author} {\bibfnamefont {D.}~\bibnamefont {Puzzuoli}},
  \bibinfo {author} {\bibfnamefont {F.}~\bibnamefont {Zhang}},\ and\ \bibinfo
  {author} {\bibfnamefont {D.~G.}\ \bibnamefont {Cory}},\ }\bibfield  {title}
  {\bibinfo {title}
  {\href{http://dx.doi.org/10.1088/1367-2630/ab4525}{Engineering effective
  Hamiltonians}},\ }\href {https://doi.org/10.1088/1367-2630/ab4525} {\bibfield
   {journal} {\bibinfo  {journal} {New J. Phys.}\ }\textbf {\bibinfo {volume}
  {21}},\ \bibinfo {pages} {103011} (\bibinfo {year} {2019})}\BibitemShut
  {NoStop}%
\bibitem [{\citenamefont {Wu}\ \emph {et~al.}(2018)\citenamefont {Wu},
  \citenamefont {Chu}, \citenamefont {Owens},\ and\ \citenamefont
  {Rabitz}}]{PhysRevA.97.042122}%
  \BibitemOpen
  \bibfield  {author} {\bibinfo {author} {\bibfnamefont {R.-B.}\ \bibnamefont
  {Wu}}, \bibinfo {author} {\bibfnamefont {B.}~\bibnamefont {Chu}}, \bibinfo
  {author} {\bibfnamefont {D.~H.}\ \bibnamefont {Owens}},\ and\ \bibinfo
  {author} {\bibfnamefont {H.}~\bibnamefont {Rabitz}},\ }\bibfield  {title}
  {\bibinfo {title} {Data-driven gradient algorithm for high-precision quantum
  control},\ }\href {https://doi.org/10.1103/PhysRevA.97.042122} {\bibfield
  {journal} {\bibinfo  {journal} {Phys. Rev. A}\ }\textbf {\bibinfo {volume}
  {97}},\ \bibinfo {pages} {042122} (\bibinfo {year} {2018})}\BibitemShut
  {NoStop}%
\bibitem [{\citenamefont {Wu}\ \emph {et~al.}(2019)\citenamefont {Wu},
  \citenamefont {Ding}, \citenamefont {Dong},\ and\ \citenamefont
  {Wang}}]{PhysRevA.99.042327}%
  \BibitemOpen
  \bibfield  {author} {\bibinfo {author} {\bibfnamefont {R.-B.}\ \bibnamefont
  {Wu}}, \bibinfo {author} {\bibfnamefont {H.}~\bibnamefont {Ding}}, \bibinfo
  {author} {\bibfnamefont {D.}~\bibnamefont {Dong}},\ and\ \bibinfo {author}
  {\bibfnamefont {X.}~\bibnamefont {Wang}},\ }\bibfield  {title} {\bibinfo
  {title} {Learning robust and high-precision quantum controls},\ }\href
  {https://doi.org/10.1103/PhysRevA.99.042327} {\bibfield  {journal} {\bibinfo
  {journal} {Phys. Rev. A}\ }\textbf {\bibinfo {volume} {99}},\ \bibinfo
  {pages} {042327} (\bibinfo {year} {2019})}\BibitemShut {NoStop}%
\bibitem [{\citenamefont {Li}\ \emph {et~al.}(2017)\citenamefont {Li},
  \citenamefont {Yang}, \citenamefont {Peng},\ and\ \citenamefont
  {Sun}}]{Li_2017}%
  \BibitemOpen
  \bibfield  {author} {\bibinfo {author} {\bibfnamefont {J.}~\bibnamefont
  {Li}}, \bibinfo {author} {\bibfnamefont {X.}~\bibnamefont {Yang}}, \bibinfo
  {author} {\bibfnamefont {X.}~\bibnamefont {Peng}},\ and\ \bibinfo {author}
  {\bibfnamefont {C.-P.}\ \bibnamefont {Sun}},\ }\bibfield  {title} {\bibinfo
  {title} {Hybrid quantum-classical approach to quantum optimal control},\
  }\bibfield  {journal} {\bibinfo  {journal} {Physical Review Letters}\
  }\textbf {\bibinfo {volume} {118}},\ \href
  {https://doi.org/10.1103/physrevlett.118.150503}
  {10.1103/physrevlett.118.150503} (\bibinfo {year} {2017})\BibitemShut
  {NoStop}%
\bibitem [{\citenamefont {Chen}\ \emph {et~al.}(2020)\citenamefont {Chen},
  \citenamefont {Yang}, \citenamefont {Arenz}, \citenamefont {Wu},
  \citenamefont {Peng}, \citenamefont {Pelczer},\ and\ \citenamefont
  {Rabitz}}]{Chen_2020}%
  \BibitemOpen
  \bibfield  {author} {\bibinfo {author} {\bibfnamefont {Q.-M.}\ \bibnamefont
  {Chen}}, \bibinfo {author} {\bibfnamefont {X.}~\bibnamefont {Yang}}, \bibinfo
  {author} {\bibfnamefont {C.}~\bibnamefont {Arenz}}, \bibinfo {author}
  {\bibfnamefont {R.-B.}\ \bibnamefont {Wu}}, \bibinfo {author} {\bibfnamefont
  {X.}~\bibnamefont {Peng}}, \bibinfo {author} {\bibfnamefont {I.}~\bibnamefont
  {Pelczer}},\ and\ \bibinfo {author} {\bibfnamefont {H.}~\bibnamefont
  {Rabitz}},\ }\bibfield  {title} {\bibinfo {title} {Combining the synergistic
  control capabilities of modeling and experiments: Illustration of finding a
  minimum-time quantum objective},\ }\bibfield  {journal} {\bibinfo  {journal}
  {Physical Review A}\ }\textbf {\bibinfo {volume} {101}},\ \href
  {https://doi.org/10.1103/physreva.101.032313} {10.1103/physreva.101.032313}
  (\bibinfo {year} {2020})\BibitemShut {NoStop}%
\bibitem [{\citenamefont {dong Yang}\ \emph {et~al.}(2020)\citenamefont {dong
  Yang}, \citenamefont {Arenz}, \citenamefont {Pelczer}, \citenamefont {Chen},
  \citenamefont {Wu}, \citenamefont {Peng},\ and\ \citenamefont
  {Rabitz}}]{Yang_2020}%
  \BibitemOpen
  \bibfield  {author} {\bibinfo {author} {\bibfnamefont {X.}~\bibnamefont {dong
  Yang}}, \bibinfo {author} {\bibfnamefont {C.}~\bibnamefont {Arenz}}, \bibinfo
  {author} {\bibfnamefont {I.}~\bibnamefont {Pelczer}}, \bibinfo {author}
  {\bibfnamefont {Q.-M.}\ \bibnamefont {Chen}}, \bibinfo {author}
  {\bibfnamefont {R.-B.}\ \bibnamefont {Wu}}, \bibinfo {author} {\bibfnamefont
  {X.}~\bibnamefont {Peng}},\ and\ \bibinfo {author} {\bibfnamefont
  {H.}~\bibnamefont {Rabitz}},\ }\bibfield  {title} {\bibinfo {title}
  {Assessing three closed-loop learning algorithms by searching for
  high-quality quantum control pulses},\ }\bibfield  {journal} {\bibinfo
  {journal} {Physical Review A}\ }\textbf {\bibinfo {volume} {102}},\ \href
  {https://doi.org/10.1103/physreva.102.062605} {10.1103/physreva.102.062605}
  (\bibinfo {year} {2020})\BibitemShut {NoStop}%
\bibitem [{\citenamefont {Dong}(2021)}]{Dong_2021}%
  \BibitemOpen
  \bibfield  {author} {\bibinfo {author} {\bibfnamefont {D.}~\bibnamefont
  {Dong}},\ }\bibfield  {title} {\bibinfo {title} {Learning control of quantum
  systems},\ }in\ \href {https://doi.org/10.1007/978-3-030-44184-5_100161}
  {\emph {\bibinfo {booktitle} {Encyclopedia of Systems and Control}}}\
  (\bibinfo  {publisher} {Springer International Publishing},\ \bibinfo {year}
  {2021})\ pp.\ \bibinfo {pages} {1090--1096}\BibitemShut {NoStop}%
\bibitem [{\citenamefont {Judson}\ and\ \citenamefont
  {Rabitz}(1992)}]{Judson_1992}%
  \BibitemOpen
  \bibfield  {author} {\bibinfo {author} {\bibfnamefont {R.~S.}\ \bibnamefont
  {Judson}}\ and\ \bibinfo {author} {\bibfnamefont {H.}~\bibnamefont
  {Rabitz}},\ }\bibfield  {title} {\bibinfo {title} {Teaching lasers to control
  molecules},\ }\href {https://doi.org/10.1103/physrevlett.68.1500} {\bibfield
  {journal} {\bibinfo  {journal} {Physical Review Letters}\ }\textbf {\bibinfo
  {volume} {68}},\ \bibinfo {pages} {1500} (\bibinfo {year}
  {1992})}\BibitemShut {NoStop}%
\bibitem [{\citenamefont {Sivak}\ \emph {et~al.}(2022)\citenamefont {Sivak},
  \citenamefont {Eickbusch}, \citenamefont {Liu}, \citenamefont {Royer},
  \citenamefont {Tsioutsios},\ and\ \citenamefont {Devoret}}]{sivak2022model}%
  \BibitemOpen
  \bibfield  {author} {\bibinfo {author} {\bibfnamefont {V.}~\bibnamefont
  {Sivak}}, \bibinfo {author} {\bibfnamefont {A.}~\bibnamefont {Eickbusch}},
  \bibinfo {author} {\bibfnamefont {H.}~\bibnamefont {Liu}}, \bibinfo {author}
  {\bibfnamefont {B.}~\bibnamefont {Royer}}, \bibinfo {author} {\bibfnamefont
  {I.}~\bibnamefont {Tsioutsios}},\ and\ \bibinfo {author} {\bibfnamefont
  {M.}~\bibnamefont {Devoret}},\ }\bibfield  {title} {\bibinfo {title}
  {Model-free quantum control with reinforcement learning},\ }\href@noop {}
  {\bibfield  {journal} {\bibinfo  {journal} {Physical Review X}\ }\textbf
  {\bibinfo {volume} {12}},\ \bibinfo {pages} {011059} (\bibinfo {year}
  {2022})}\BibitemShut {NoStop}%
\bibitem [{\citenamefont {Baum}\ \emph {et~al.}(2021)\citenamefont {Baum},
  \citenamefont {Amico}, \citenamefont {Howell}, \citenamefont {Hush},
  \citenamefont {Liuzzi}, \citenamefont {Mundada}, \citenamefont {Merkh},
  \citenamefont {Carvalho},\ and\ \citenamefont
  {Biercuk}}]{baum2021experimental}%
  \BibitemOpen
  \bibfield  {author} {\bibinfo {author} {\bibfnamefont {Y.}~\bibnamefont
  {Baum}}, \bibinfo {author} {\bibfnamefont {M.}~\bibnamefont {Amico}},
  \bibinfo {author} {\bibfnamefont {S.}~\bibnamefont {Howell}}, \bibinfo
  {author} {\bibfnamefont {M.}~\bibnamefont {Hush}}, \bibinfo {author}
  {\bibfnamefont {M.}~\bibnamefont {Liuzzi}}, \bibinfo {author} {\bibfnamefont
  {P.}~\bibnamefont {Mundada}}, \bibinfo {author} {\bibfnamefont
  {T.}~\bibnamefont {Merkh}}, \bibinfo {author} {\bibfnamefont {A.~R.}\
  \bibnamefont {Carvalho}},\ and\ \bibinfo {author} {\bibfnamefont {M.~J.}\
  \bibnamefont {Biercuk}},\ }\bibfield  {title} {\bibinfo {title} {Experimental
  deep reinforcement learning for error-robust gate-set design on a
  superconducting quantum computer},\ }\href@noop {} {\bibfield  {journal}
  {\bibinfo  {journal} {PRX Quantum}\ }\textbf {\bibinfo {volume} {2}},\
  \bibinfo {pages} {040324} (\bibinfo {year} {2021})}\BibitemShut {NoStop}%
\bibitem [{\citenamefont {Niu}\ \emph {et~al.}(2019)\citenamefont {Niu},
  \citenamefont {Boixo}, \citenamefont {Smelyanskiy},\ and\ \citenamefont
  {Neven}}]{niu2019universal}%
  \BibitemOpen
  \bibfield  {author} {\bibinfo {author} {\bibfnamefont {M.~Y.}\ \bibnamefont
  {Niu}}, \bibinfo {author} {\bibfnamefont {S.}~\bibnamefont {Boixo}}, \bibinfo
  {author} {\bibfnamefont {V.~N.}\ \bibnamefont {Smelyanskiy}},\ and\ \bibinfo
  {author} {\bibfnamefont {H.}~\bibnamefont {Neven}},\ }\bibfield  {title}
  {\bibinfo {title} {\href{https://doi.org/10.1038/s41534-019-0141-3}{Universal
  quantum control through deep reinforcement learning}},\ }\bibfield  {journal}
  {\bibinfo  {journal} {npj Quantum Inf.}\ }\textbf {\bibinfo {volume} {5}},\
  \href {https://doi.org/10.1038/s41534-019-0141-3} {10.1038/s41534-019-0141-3}
  (\bibinfo {year} {2019})\BibitemShut {NoStop}%
\bibitem [{\citenamefont {Erdman}\ and\ \citenamefont
  {No{\'e}}(2022)}]{erdman2022driving}%
  \BibitemOpen
  \bibfield  {author} {\bibinfo {author} {\bibfnamefont {P.~A.}\ \bibnamefont
  {Erdman}}\ and\ \bibinfo {author} {\bibfnamefont {F.}~\bibnamefont
  {No{\'e}}},\ }\bibfield  {title} {\bibinfo {title} {Driving black-box quantum
  thermal machines with optimal power/efficiency trade-offs using reinforcement
  learning},\ }\href@noop {} {\bibfield  {journal} {\bibinfo  {journal} {arXiv
  preprint arXiv:2204.04785}\ } (\bibinfo {year} {2022})}\BibitemShut {NoStop}%
\bibitem [{\citenamefont {Matthews}\ \emph {et~al.}(2009)\citenamefont
  {Matthews}, \citenamefont {Politi}, \citenamefont {Stefanov},\ and\
  \citenamefont {O'Brien}}]{matthews_manipulation_2009}%
  \BibitemOpen
  \bibfield  {author} {\bibinfo {author} {\bibfnamefont {J.~C.~F.}\
  \bibnamefont {Matthews}}, \bibinfo {author} {\bibfnamefont {A.}~\bibnamefont
  {Politi}}, \bibinfo {author} {\bibfnamefont {A.}~\bibnamefont {Stefanov}},\
  and\ \bibinfo {author} {\bibfnamefont {J.~L.}\ \bibnamefont {O'Brien}},\
  }\bibfield  {title} {\bibinfo {title} {Manipulation of multiphoton
  entanglement in waveguide quantum circuits},\ }\href
  {https://doi.org/10.1038/nphoton.2009.93} {\bibfield  {journal} {\bibinfo
  {journal} {Nature Photonics}\ }\textbf {\bibinfo {volume} {3}},\ \bibinfo
  {pages} {346} (\bibinfo {year} {2009})}\BibitemShut {NoStop}%
\bibitem [{\citenamefont {Erdman}\ and\ \citenamefont
  {No{\'{e}}}(2022)}]{Erdman_2022}%
  \BibitemOpen
  \bibfield  {author} {\bibinfo {author} {\bibfnamefont {P.~A.}\ \bibnamefont
  {Erdman}}\ and\ \bibinfo {author} {\bibfnamefont {F.}~\bibnamefont
  {No{\'{e}}}},\ }\bibfield  {title} {\bibinfo {title} {Identifying optimal
  cycles in quantum thermal machines with reinforcement-learning},\ }\bibfield
  {journal} {\bibinfo  {journal} {npj Quantum Information}\ }\textbf {\bibinfo
  {volume} {8}},\ \href {https://doi.org/10.1038/s41534-021-00512-0}
  {10.1038/s41534-021-00512-0} (\bibinfo {year} {2022})\BibitemShut {NoStop}%
\bibitem [{\citenamefont {Fanchini}\ \emph {et~al.}(2021)\citenamefont
  {Fanchini}, \citenamefont {Karpat}, \citenamefont {Rossatto}, \citenamefont
  {Norambuena},\ and\ \citenamefont {Coto}}]{fanchini2021estimating}%
  \BibitemOpen
  \bibfield  {author} {\bibinfo {author} {\bibfnamefont {F.~F.}\ \bibnamefont
  {Fanchini}}, \bibinfo {author} {\bibfnamefont {G.}~\bibnamefont {Karpat}},
  \bibinfo {author} {\bibfnamefont {D.~Z.}\ \bibnamefont {Rossatto}}, \bibinfo
  {author} {\bibfnamefont {A.}~\bibnamefont {Norambuena}},\ and\ \bibinfo
  {author} {\bibfnamefont {R.}~\bibnamefont {Coto}},\ }\bibfield  {title}
  {\bibinfo {title} {Estimating the degree of non-markovianity using machine
  learning},\ }\href@noop {} {\bibfield  {journal} {\bibinfo  {journal}
  {Physical Review A}\ }\textbf {\bibinfo {volume} {103}},\ \bibinfo {pages}
  {022425} (\bibinfo {year} {2021})}\BibitemShut {NoStop}%
\bibitem [{\citenamefont {Flurin}\ \emph {et~al.}(2020)\citenamefont {Flurin},
  \citenamefont {Martin}, \citenamefont {Hacohen-Gourgy},\ and\ \citenamefont
  {Siddiqi}}]{flurin2020using}%
  \BibitemOpen
  \bibfield  {author} {\bibinfo {author} {\bibfnamefont {E.}~\bibnamefont
  {Flurin}}, \bibinfo {author} {\bibfnamefont {L.~S.}\ \bibnamefont {Martin}},
  \bibinfo {author} {\bibfnamefont {S.}~\bibnamefont {Hacohen-Gourgy}},\ and\
  \bibinfo {author} {\bibfnamefont {I.}~\bibnamefont {Siddiqi}},\ }\bibfield
  {title} {\bibinfo {title} {Using a recurrent neural network to reconstruct
  quantum dynamics of a superconducting qubit from physical observations},\
  }\href@noop {} {\bibfield  {journal} {\bibinfo  {journal} {Physical Review
  X}\ }\textbf {\bibinfo {volume} {10}},\ \bibinfo {pages} {011006} (\bibinfo
  {year} {2020})}\BibitemShut {NoStop}%
\bibitem [{\citenamefont {Papi{\v{c}}}\ and\ \citenamefont
  {de~Vega}(2022)}]{papivc2022neural}%
  \BibitemOpen
  \bibfield  {author} {\bibinfo {author} {\bibfnamefont {M.}~\bibnamefont
  {Papi{\v{c}}}}\ and\ \bibinfo {author} {\bibfnamefont {I.}~\bibnamefont
  {de~Vega}},\ }\bibfield  {title} {\bibinfo {title} {Neural-network-based
  qubit-environment characterization},\ }\href@noop {} {\bibfield  {journal}
  {\bibinfo  {journal} {Physical Review A}\ }\textbf {\bibinfo {volume}
  {105}},\ \bibinfo {pages} {022605} (\bibinfo {year} {2022})}\BibitemShut
  {NoStop}%
\bibitem [{\citenamefont {Wise}\ \emph {et~al.}(2021)\citenamefont {Wise},
  \citenamefont {Morton},\ and\ \citenamefont {Dhomkar}}]{wise2021using}%
  \BibitemOpen
  \bibfield  {author} {\bibinfo {author} {\bibfnamefont {D.~F.}\ \bibnamefont
  {Wise}}, \bibinfo {author} {\bibfnamefont {J.~J.}\ \bibnamefont {Morton}},\
  and\ \bibinfo {author} {\bibfnamefont {S.}~\bibnamefont {Dhomkar}},\
  }\bibfield  {title} {\bibinfo {title} {Using deep learning to understand and
  mitigate the qubit noise environment},\ }\href@noop {} {\bibfield  {journal}
  {\bibinfo  {journal} {PRX Quantum}\ }\textbf {\bibinfo {volume} {2}},\
  \bibinfo {pages} {010316} (\bibinfo {year} {2021})}\BibitemShut {NoStop}%
\bibitem [{\citenamefont {Palmieri}\ \emph {et~al.}(2021)\citenamefont
  {Palmieri}, \citenamefont {Bianchi}, \citenamefont {Paris},\ and\
  \citenamefont {Benedetti}}]{palmieri2021multiclass}%
  \BibitemOpen
  \bibfield  {author} {\bibinfo {author} {\bibfnamefont {A.~M.}\ \bibnamefont
  {Palmieri}}, \bibinfo {author} {\bibfnamefont {F.}~\bibnamefont {Bianchi}},
  \bibinfo {author} {\bibfnamefont {M.~G.}\ \bibnamefont {Paris}},\ and\
  \bibinfo {author} {\bibfnamefont {C.}~\bibnamefont {Benedetti}},\ }\bibfield
  {title} {\bibinfo {title} {Multiclass classification of dephasing channels},\
  }\href@noop {} {\bibfield  {journal} {\bibinfo  {journal} {Physical Review
  A}\ }\textbf {\bibinfo {volume} {104}},\ \bibinfo {pages} {052412} (\bibinfo
  {year} {2021})}\BibitemShut {NoStop}%
\bibitem [{\citenamefont {Ostaszewski}\ \emph {et~al.}(2019)\citenamefont
  {Ostaszewski}, \citenamefont {Miszczak}, \citenamefont {Banchi},\ and\
  \citenamefont {Sadowski}}]{ostaszewski2019approximation}%
  \BibitemOpen
  \bibfield  {author} {\bibinfo {author} {\bibfnamefont {M.}~\bibnamefont
  {Ostaszewski}}, \bibinfo {author} {\bibfnamefont {J.}~\bibnamefont
  {Miszczak}}, \bibinfo {author} {\bibfnamefont {L.}~\bibnamefont {Banchi}},\
  and\ \bibinfo {author} {\bibfnamefont {P.}~\bibnamefont {Sadowski}},\
  }\bibfield  {title} {\bibinfo {title} {Approximation of quantum control
  correction scheme using deep neural networks},\ }\href@noop {} {\bibfield
  {journal} {\bibinfo  {journal} {Quantum Information Processing}\ }\textbf
  {\bibinfo {volume} {18}},\ \bibinfo {pages} {1} (\bibinfo {year}
  {2019})}\BibitemShut {NoStop}%
\bibitem [{\citenamefont {Khait}\ \emph {et~al.}(2022)\citenamefont {Khait},
  \citenamefont {Carrasquilla},\ and\ \citenamefont
  {Segal}}]{khait2022optimal}%
  \BibitemOpen
  \bibfield  {author} {\bibinfo {author} {\bibfnamefont {I.}~\bibnamefont
  {Khait}}, \bibinfo {author} {\bibfnamefont {J.}~\bibnamefont
  {Carrasquilla}},\ and\ \bibinfo {author} {\bibfnamefont {D.}~\bibnamefont
  {Segal}},\ }\bibfield  {title} {\bibinfo {title} {Optimal control of quantum
  thermal machines using machine learning},\ }\href@noop {} {\bibfield
  {journal} {\bibinfo  {journal} {Physical Review Research}\ }\textbf {\bibinfo
  {volume} {4}},\ \bibinfo {pages} {L012029} (\bibinfo {year}
  {2022})}\BibitemShut {NoStop}%
\bibitem [{\citenamefont {Zeng}\ \emph {et~al.}(2020)\citenamefont {Zeng},
  \citenamefont {Shen}, \citenamefont {Hou}, \citenamefont {Gebremariam},\ and\
  \citenamefont {Li}}]{zeng2020quantum}%
  \BibitemOpen
  \bibfield  {author} {\bibinfo {author} {\bibfnamefont {Y.}~\bibnamefont
  {Zeng}}, \bibinfo {author} {\bibfnamefont {J.}~\bibnamefont {Shen}}, \bibinfo
  {author} {\bibfnamefont {S.}~\bibnamefont {Hou}}, \bibinfo {author}
  {\bibfnamefont {T.}~\bibnamefont {Gebremariam}},\ and\ \bibinfo {author}
  {\bibfnamefont {C.}~\bibnamefont {Li}},\ }\bibfield  {title} {\bibinfo
  {title} {Quantum control based on machine learning in an open quantum
  system},\ }\href@noop {} {\bibfield  {journal} {\bibinfo  {journal} {Physics
  Letters A}\ }\textbf {\bibinfo {volume} {384}},\ \bibinfo {pages} {126886}
  (\bibinfo {year} {2020})}\BibitemShut {NoStop}%
\bibitem [{\citenamefont {Bausch}\ and\ \citenamefont
  {Leditzky}(2020)}]{bausch2020quantum}%
  \BibitemOpen
  \bibfield  {author} {\bibinfo {author} {\bibfnamefont {J.}~\bibnamefont
  {Bausch}}\ and\ \bibinfo {author} {\bibfnamefont {F.}~\bibnamefont
  {Leditzky}},\ }\bibfield  {title} {\bibinfo {title} {Quantum codes from
  neural networks},\ }\href@noop {} {\bibfield  {journal} {\bibinfo  {journal}
  {New Journal of Physics}\ }\textbf {\bibinfo {volume} {22}},\ \bibinfo
  {pages} {023005} (\bibinfo {year} {2020})}\BibitemShut {NoStop}%
\bibitem [{\citenamefont {Baireuther}\ \emph {et~al.}(2018)\citenamefont
  {Baireuther}, \citenamefont {O'Brien}, \citenamefont {Tarasinski},\ and\
  \citenamefont {Beenakker}}]{baireuther2018machine}%
  \BibitemOpen
  \bibfield  {author} {\bibinfo {author} {\bibfnamefont {P.}~\bibnamefont
  {Baireuther}}, \bibinfo {author} {\bibfnamefont {T.~E.}\ \bibnamefont
  {O'Brien}}, \bibinfo {author} {\bibfnamefont {B.}~\bibnamefont
  {Tarasinski}},\ and\ \bibinfo {author} {\bibfnamefont {C.~W.}\ \bibnamefont
  {Beenakker}},\ }\bibfield  {title} {\bibinfo {title}
  {Machine-learning-assisted correction of correlated qubit errors in a
  topological code},\ }\href@noop {} {\bibfield  {journal} {\bibinfo  {journal}
  {Quantum}\ }\textbf {\bibinfo {volume} {2}},\ \bibinfo {pages} {48} (\bibinfo
  {year} {2018})}\BibitemShut {NoStop}%
\bibitem [{\citenamefont {Lennon}\ \emph {et~al.}(2019)\citenamefont {Lennon},
  \citenamefont {Moon}, \citenamefont {Camenzind}, \citenamefont {Yu},
  \citenamefont {Zumb{\"u}hl}, \citenamefont {Briggs}, \citenamefont {Osborne},
  \citenamefont {Laird},\ and\ \citenamefont {Ares}}]{lennon2019efficiently}%
  \BibitemOpen
  \bibfield  {author} {\bibinfo {author} {\bibfnamefont {D.~T.}\ \bibnamefont
  {Lennon}}, \bibinfo {author} {\bibfnamefont {H.}~\bibnamefont {Moon}},
  \bibinfo {author} {\bibfnamefont {L.~C.}\ \bibnamefont {Camenzind}}, \bibinfo
  {author} {\bibfnamefont {L.}~\bibnamefont {Yu}}, \bibinfo {author}
  {\bibfnamefont {D.~M.}\ \bibnamefont {Zumb{\"u}hl}}, \bibinfo {author}
  {\bibfnamefont {G.~A.~D.}\ \bibnamefont {Briggs}}, \bibinfo {author}
  {\bibfnamefont {M.~A.}\ \bibnamefont {Osborne}}, \bibinfo {author}
  {\bibfnamefont {E.~A.}\ \bibnamefont {Laird}},\ and\ \bibinfo {author}
  {\bibfnamefont {N.}~\bibnamefont {Ares}},\ }\bibfield  {title} {\bibinfo
  {title} {Efficiently measuring a quantum device using machine learning},\
  }\href@noop {} {\bibfield  {journal} {\bibinfo  {journal} {npj Quantum
  Information}\ }\textbf {\bibinfo {volume} {5}},\ \bibinfo {pages} {1}
  (\bibinfo {year} {2019})}\BibitemShut {NoStop}%
\bibitem [{\citenamefont {Tranter}\ \emph {et~al.}(2018)\citenamefont
  {Tranter}, \citenamefont {Slatyer}, \citenamefont {Hush}, \citenamefont
  {Leung}, \citenamefont {Everett}, \citenamefont {Paul}, \citenamefont
  {Vernaz-Gris}, \citenamefont {Lam}, \citenamefont {Buchler},\ and\
  \citenamefont {Campbell}}]{10.1038/s41467-018-06847-1}%
  \BibitemOpen
  \bibfield  {author} {\bibinfo {author} {\bibfnamefont {A.~D.}\ \bibnamefont
  {Tranter}}, \bibinfo {author} {\bibfnamefont {H.~J.}\ \bibnamefont
  {Slatyer}}, \bibinfo {author} {\bibfnamefont {M.~R.}\ \bibnamefont {Hush}},
  \bibinfo {author} {\bibfnamefont {A.~C.}\ \bibnamefont {Leung}}, \bibinfo
  {author} {\bibfnamefont {J.~L.}\ \bibnamefont {Everett}}, \bibinfo {author}
  {\bibfnamefont {K.~V.}\ \bibnamefont {Paul}}, \bibinfo {author}
  {\bibfnamefont {P.}~\bibnamefont {Vernaz-Gris}}, \bibinfo {author}
  {\bibfnamefont {P.~K.}\ \bibnamefont {Lam}}, \bibinfo {author} {\bibfnamefont
  {B.~C.}\ \bibnamefont {Buchler}},\ and\ \bibinfo {author} {\bibfnamefont
  {G.~T.}\ \bibnamefont {Campbell}},\ }\bibfield  {title} {\bibinfo {title}
  {Multiparameter optimisation of a magneto-optical trap using deep learning},\
  }\href {https://doi.org/10.1038/s41467-018-06847-1} {\bibfield  {journal}
  {\bibinfo  {journal} {Nature Communications}\ }\textbf {\bibinfo {volume}
  {9}},\ \bibinfo {pages} {4360} (\bibinfo {year} {2018})}\BibitemShut
  {NoStop}%
\bibitem [{\citenamefont {Cimini}\ \emph {et~al.}(2021)\citenamefont {Cimini},
  \citenamefont {Polino}, \citenamefont {Valeri}, \citenamefont {Gianani},
  \citenamefont {Spagnolo}, \citenamefont {Corrielli}, \citenamefont {Crespi},
  \citenamefont {Osellame}, \citenamefont {Barbieri},\ and\ \citenamefont
  {Sciarrino}}]{10.1103/physrevapplied.15.044003}%
  \BibitemOpen
  \bibfield  {author} {\bibinfo {author} {\bibfnamefont {V.}~\bibnamefont
  {Cimini}}, \bibinfo {author} {\bibfnamefont {E.}~\bibnamefont {Polino}},
  \bibinfo {author} {\bibfnamefont {M.}~\bibnamefont {Valeri}}, \bibinfo
  {author} {\bibfnamefont {I.}~\bibnamefont {Gianani}}, \bibinfo {author}
  {\bibfnamefont {N.}~\bibnamefont {Spagnolo}}, \bibinfo {author}
  {\bibfnamefont {G.}~\bibnamefont {Corrielli}}, \bibinfo {author}
  {\bibfnamefont {A.}~\bibnamefont {Crespi}}, \bibinfo {author} {\bibfnamefont
  {R.}~\bibnamefont {Osellame}}, \bibinfo {author} {\bibfnamefont
  {M.}~\bibnamefont {Barbieri}},\ and\ \bibinfo {author} {\bibfnamefont
  {F.}~\bibnamefont {Sciarrino}},\ }\bibfield  {title} {\bibinfo {title}
  {{Calibration of Multiparameter Sensors via Machine Learning at the
  Single-Photon Level}},\ }\href
  {https://doi.org/10.1103/physrevapplied.15.044003} {\bibfield  {journal}
  {\bibinfo  {journal} {Physical Review Applied}\ }\textbf {\bibinfo {volume}
  {15}},\ \bibinfo {pages} {044003} (\bibinfo {year} {2021})},\ \Eprint
  {https://arxiv.org/abs/2009.07122} {2009.07122} \BibitemShut {NoStop}%
\bibitem [{\citenamefont {Wiebe}\ \emph {et~al.}(2014)\citenamefont {Wiebe},
  \citenamefont {Granade}, \citenamefont {Ferrie},\ and\ \citenamefont
  {Cory}}]{wiebe2014quantum}%
  \BibitemOpen
  \bibfield  {author} {\bibinfo {author} {\bibfnamefont {N.}~\bibnamefont
  {Wiebe}}, \bibinfo {author} {\bibfnamefont {C.}~\bibnamefont {Granade}},
  \bibinfo {author} {\bibfnamefont {C.}~\bibnamefont {Ferrie}},\ and\ \bibinfo
  {author} {\bibfnamefont {D.}~\bibnamefont {Cory}},\ }\bibfield  {title}
  {\bibinfo {title} {\href{https://doi.org/10.1103/physreva.89.042314}{Quantum
  Hamiltonian learning using imperfect quantum resources}},\ }\href
  {https://doi.org/10.1103/physreva.89.042314} {\bibfield  {journal} {\bibinfo
  {journal} {Phys. Rev. A}\ }\textbf {\bibinfo {volume} {89}},\ \bibinfo
  {pages} {042314} (\bibinfo {year} {2014})}\BibitemShut {NoStop}%
\bibitem [{\citenamefont {Wang}\ \emph {et~al.}(2017)\citenamefont {Wang} \emph
  {et~al.}}]{wang2017experimental}%
  \BibitemOpen
  \bibfield  {author} {\bibinfo {author} {\bibfnamefont {J.}~\bibnamefont
  {Wang}} \emph {et~al.},\ }\bibfield  {title} {\bibinfo {title}
  {\href{https://doi.org/10.1038/nphys4074}{Experimental quantum Hamiltonian
  learning}},\ }\href {https://doi.org/10.1038/nphys4074} {\bibfield  {journal}
  {\bibinfo  {journal} {Nat. Phys.}\ }\textbf {\bibinfo {volume} {13}},\
  \bibinfo {pages} {551} (\bibinfo {year} {2017})}\BibitemShut {NoStop}%
\bibitem [{GB_(2006)}]{GB_book}%
  \BibitemOpen
  \href {https://doi.org/10.1007/1-84628-403-1} {\emph {\bibinfo {title}
  {Practical Grey-box Process Identification}}}\ (\bibinfo  {publisher}
  {Springer London},\ \bibinfo {year} {2006})\BibitemShut {NoStop}%
\bibitem [{\citenamefont {Youssry}\ \emph
  {et~al.}(2020{\natexlab{a}})\citenamefont {Youssry}, \citenamefont {Chapman},
  \citenamefont {Peruzzo}, \citenamefont {Ferrie},\ and\ \citenamefont
  {Tomamichel}}]{youssry2020modeling}%
  \BibitemOpen
  \bibfield  {author} {\bibinfo {author} {\bibfnamefont {A.}~\bibnamefont
  {Youssry}}, \bibinfo {author} {\bibfnamefont {R.~J.}\ \bibnamefont
  {Chapman}}, \bibinfo {author} {\bibfnamefont {A.}~\bibnamefont {Peruzzo}},
  \bibinfo {author} {\bibfnamefont {C.}~\bibnamefont {Ferrie}},\ and\ \bibinfo
  {author} {\bibfnamefont {M.}~\bibnamefont {Tomamichel}},\ }\bibfield  {title}
  {\bibinfo {title} {{Modeling and control of a reconfigurable photonic circuit
  using deep learning}},\ }\href@noop {} {\bibfield  {journal} {\bibinfo
  {journal} {Quantum Science and Technology}\ }\textbf {\bibinfo {volume}
  {5}},\ \bibinfo {pages} {025001} (\bibinfo {year}
  {2020}{\natexlab{a}})}\BibitemShut {NoStop}%
\bibitem [{\citenamefont {Youssry}\ \emph
  {et~al.}(2020{\natexlab{b}})\citenamefont {Youssry}, \citenamefont
  {Paz-Silva},\ and\ \citenamefont {Ferrie}}]{youssry2020characterization}%
  \BibitemOpen
  \bibfield  {author} {\bibinfo {author} {\bibfnamefont {A.}~\bibnamefont
  {Youssry}}, \bibinfo {author} {\bibfnamefont {G.~A.}\ \bibnamefont
  {Paz-Silva}},\ and\ \bibinfo {author} {\bibfnamefont {C.}~\bibnamefont
  {Ferrie}},\ }\bibfield  {title} {\bibinfo {title} {{Characterization and
  control of open quantum systems beyond quantum noise spectroscopy}},\
  }\href@noop {} {\bibfield  {journal} {\bibinfo  {journal} {npj Quantum
  Information}\ }\textbf {\bibinfo {volume} {6}},\ \bibinfo {pages} {1}
  (\bibinfo {year} {2020}{\natexlab{b}})}\BibitemShut {NoStop}%
\bibitem [{\citenamefont {Youssry}\ and\ \citenamefont
  {Nurdin}(2023)}]{youssry2022Multi}%
  \BibitemOpen
  \bibfield  {author} {\bibinfo {author} {\bibfnamefont {A.}~\bibnamefont
  {Youssry}}\ and\ \bibinfo {author} {\bibfnamefont {H.~I.}\ \bibnamefont
  {Nurdin}},\ }\bibfield  {title} {\bibinfo {title} {{Multi-axis control of a
  qubit in the presence of unknown non-Markovian quantum noise}},\ }\href
  {https://doi.org/10.1088/2058-9565/aca711} {\bibfield  {journal} {\bibinfo
  {journal} {Quantum Science and Technology}\ }\textbf {\bibinfo {volume}
  {8}},\ \bibinfo {pages} {015018} (\bibinfo {year} {2023})},\ \Eprint
  {https://arxiv.org/abs/2208.03058} {2208.03058} \BibitemShut {NoStop}%
\bibitem [{\citenamefont {Youssry}\ \emph {et~al.}(2023)\citenamefont
  {Youssry}, \citenamefont {Paz-Silva},\ and\ \citenamefont
  {Ferrie}}]{youssry2021noise}%
  \BibitemOpen
  \bibfield  {author} {\bibinfo {author} {\bibfnamefont {A.}~\bibnamefont
  {Youssry}}, \bibinfo {author} {\bibfnamefont {G.~A.}\ \bibnamefont
  {Paz-Silva}},\ and\ \bibinfo {author} {\bibfnamefont {C.}~\bibnamefont
  {Ferrie}},\ }\bibfield  {title} {\bibinfo {title} {Noise detection with
  spectator qubits and quantum feature engineering},\ }\href
  {https://doi.org/10.1088/1367-2630/ace2e4} {\bibfield  {journal} {\bibinfo
  {journal} {New Journal of Physics}\ }\textbf {\bibinfo {volume} {25}},\
  \bibinfo {pages} {073004} (\bibinfo {year} {2023})}\BibitemShut {NoStop}%
\bibitem [{\citenamefont {Perrier}\ \emph {et~al.}(2020)\citenamefont
  {Perrier}, \citenamefont {Tao},\ and\ \citenamefont
  {Ferrie}}]{perrier2020quantum}%
  \BibitemOpen
  \bibfield  {author} {\bibinfo {author} {\bibfnamefont {E.}~\bibnamefont
  {Perrier}}, \bibinfo {author} {\bibfnamefont {D.}~\bibnamefont {Tao}},\ and\
  \bibinfo {author} {\bibfnamefont {C.}~\bibnamefont {Ferrie}},\ }\bibfield
  {title} {\bibinfo {title} {Quantum geometric machine learning for quantum
  circuits and control},\ }\href@noop {} {\bibfield  {journal} {\bibinfo
  {journal} {New Journal of Physics}\ }\textbf {\bibinfo {volume} {22}},\
  \bibinfo {pages} {103056} (\bibinfo {year} {2020})}\BibitemShut {NoStop}%
\bibitem [{\citenamefont {Genois}\ \emph {et~al.}(2021)\citenamefont {Genois},
  \citenamefont {Gross}, \citenamefont {Paolo}, \citenamefont {Stevenson},
  \citenamefont {Koolstra}, \citenamefont {Hashim}, \citenamefont {Siddiqi},\
  and\ \citenamefont {Blais}}]{10.1103/prxquantum.2.040355}%
  \BibitemOpen
  \bibfield  {author} {\bibinfo {author} {\bibfnamefont {{\'E}.}~\bibnamefont
  {Genois}}, \bibinfo {author} {\bibfnamefont {J.~A.}\ \bibnamefont {Gross}},
  \bibinfo {author} {\bibfnamefont {A.~D.}\ \bibnamefont {Paolo}}, \bibinfo
  {author} {\bibfnamefont {N.~J.}\ \bibnamefont {Stevenson}}, \bibinfo {author}
  {\bibfnamefont {G.}~\bibnamefont {Koolstra}}, \bibinfo {author}
  {\bibfnamefont {A.}~\bibnamefont {Hashim}}, \bibinfo {author} {\bibfnamefont
  {I.}~\bibnamefont {Siddiqi}},\ and\ \bibinfo {author} {\bibfnamefont
  {A.}~\bibnamefont {Blais}},\ }\bibfield  {title} {\bibinfo {title}
  {{Quantum-Tailored Machine-Learning Characterization of a Superconducting
  Qubit}},\ }\href {https://doi.org/10.1103/prxquantum.2.040355} {\bibfield
  {journal} {\bibinfo  {journal} {PRX Quantum}\ }\textbf {\bibinfo {volume}
  {2}},\ \bibinfo {pages} {040355} (\bibinfo {year} {2021})},\ \Eprint
  {https://arxiv.org/abs/2106.13126} {2106.13126} \BibitemShut {NoStop}%
\bibitem [{\citenamefont {Peruzzo}\ \emph {et~al.}(2010)\citenamefont
  {Peruzzo}, \citenamefont {Lobino}, \citenamefont {Matthews}, \citenamefont
  {Matsuda}, \citenamefont {Politi}, \citenamefont {Poulios}, \citenamefont
  {Zhou}, \citenamefont {Lahini}, \citenamefont {Ismail}, \citenamefont
  {Wörhoff}, \citenamefont {Bromberg}, \citenamefont {Silberberg},
  \citenamefont {Thompson},\ and\ \citenamefont
  {OBrien}}]{Peruzzo.2010.Science.10.1126/science.1193515j8}%
  \BibitemOpen
  \bibfield  {author} {\bibinfo {author} {\bibfnamefont {A.}~\bibnamefont
  {Peruzzo}}, \bibinfo {author} {\bibfnamefont {M.}~\bibnamefont {Lobino}},
  \bibinfo {author} {\bibfnamefont {J.~C.~F.}\ \bibnamefont {Matthews}},
  \bibinfo {author} {\bibfnamefont {N.}~\bibnamefont {Matsuda}}, \bibinfo
  {author} {\bibfnamefont {A.}~\bibnamefont {Politi}}, \bibinfo {author}
  {\bibfnamefont {K.}~\bibnamefont {Poulios}}, \bibinfo {author} {\bibfnamefont
  {X.-Q.}\ \bibnamefont {Zhou}}, \bibinfo {author} {\bibfnamefont
  {Y.}~\bibnamefont {Lahini}}, \bibinfo {author} {\bibfnamefont
  {N.}~\bibnamefont {Ismail}}, \bibinfo {author} {\bibfnamefont
  {K.}~\bibnamefont {Wörhoff}}, \bibinfo {author} {\bibfnamefont
  {Y.}~\bibnamefont {Bromberg}}, \bibinfo {author} {\bibfnamefont
  {Y.}~\bibnamefont {Silberberg}}, \bibinfo {author} {\bibfnamefont {M.~G.}\
  \bibnamefont {Thompson}},\ and\ \bibinfo {author} {\bibfnamefont {J.~L.}\
  \bibnamefont {OBrien}},\ }\bibfield  {title} {\bibinfo {title} {{Quantum
  Walks of Correlated Photons}},\ }\href
  {https://doi.org/10.1126/science.1193515} {\bibfield  {journal} {\bibinfo
  {journal} {Science}\ }\textbf {\bibinfo {volume} {329}},\ \bibinfo {pages}
  {1500} (\bibinfo {year} {2010})},\ \Eprint {https://arxiv.org/abs/1006.4764}
  {1006.4764} \BibitemShut {NoStop}%
\bibitem [{\citenamefont {Tieleman}\ and\ \citenamefont
  {Hinton}(2012)}]{rmsprop}%
  \BibitemOpen
  \bibfield  {author} {\bibinfo {author} {\bibfnamefont {T.}~\bibnamefont
  {Tieleman}}\ and\ \bibinfo {author} {\bibfnamefont {G.}~\bibnamefont
  {Hinton}},\ }\bibfield  {title} {\bibinfo {title} {Lecture 6.5-rmsprop:
  Divide the gradient by a running average of its recent magnitude},\
  }\href@noop {} {\bibfield  {journal} {\bibinfo  {journal} {COURSERA: Neural
  networks for machine learning}\ }\textbf {\bibinfo {volume} {4}},\ \bibinfo
  {pages} {26} (\bibinfo {year} {2012})}\BibitemShut {NoStop}%
\bibitem [{\citenamefont {Kingma}\ and\ \citenamefont {Ba}(2015)}]{Adam}%
  \BibitemOpen
  \bibfield  {author} {\bibinfo {author} {\bibfnamefont {D.~P.}\ \bibnamefont
  {Kingma}}\ and\ \bibinfo {author} {\bibfnamefont {J.}~\bibnamefont {Ba}},\
  }\bibfield  {title} {\bibinfo {title} {Adam: {A} method for stochastic
  optimization},\ }in\ \href {http://arxiv.org/abs/1412.6980} {\emph {\bibinfo
  {booktitle} {3rd International Conference on Learning Representations, {ICLR}
  2015, San Diego, CA, USA, May 7-9, 2015, Conference Track Proceedings}}},\
  \bibinfo {editor} {edited by\ \bibinfo {editor} {\bibfnamefont
  {Y.}~\bibnamefont {Bengio}}\ and\ \bibinfo {editor} {\bibfnamefont
  {Y.}~\bibnamefont {LeCun}}}\ (\bibinfo {year} {2015})\BibitemShut {NoStop}%
\bibitem [{\citenamefont {Prencipe}\ and\ \citenamefont
  {Gallo}(2023)}]{10011218}%
  \BibitemOpen
  \bibfield  {author} {\bibinfo {author} {\bibfnamefont {A.}~\bibnamefont
  {Prencipe}}\ and\ \bibinfo {author} {\bibfnamefont {K.}~\bibnamefont
  {Gallo}},\ }\bibfield  {title} {\bibinfo {title} {Electro- and thermo-optics
  response of x-cut thin film linbo3 waveguides},\ }\href
  {https://doi.org/10.1109/JQE.2023.3234986} {\bibfield  {journal} {\bibinfo
  {journal} {IEEE Journal of Quantum Electronics}\ }\textbf {\bibinfo {volume}
  {59}},\ \bibinfo {pages} {1} (\bibinfo {year} {2023})}\BibitemShut {NoStop}%
\bibitem [{\citenamefont {Abadi}\ \emph {et~al.}(2015)\citenamefont {Abadi},
  \citenamefont {Agarwal}, \citenamefont {Barham}, \citenamefont {Brevdo},
  \citenamefont {Chen}, \citenamefont {Citro}, \citenamefont {Corrado},
  \citenamefont {Davis}, \citenamefont {Dean}, \citenamefont {Devin},
  \citenamefont {Ghemawat}, \citenamefont {Goodfellow}, \citenamefont {Harp},
  \citenamefont {Irving}, \citenamefont {Isard}, \citenamefont {Jia},
  \citenamefont {Jozefowicz}, \citenamefont {Kaiser}, \citenamefont {Kudlur},
  \citenamefont {Levenberg}, \citenamefont {Man\'{e}}, \citenamefont {Monga},
  \citenamefont {Moore}, \citenamefont {Murray}, \citenamefont {Olah},
  \citenamefont {Schuster}, \citenamefont {Shlens}, \citenamefont {Steiner},
  \citenamefont {Sutskever}, \citenamefont {Talwar}, \citenamefont {Tucker},
  \citenamefont {Vanhoucke}, \citenamefont {Vasudevan}, \citenamefont
  {Vi\'{e}gas}, \citenamefont {Vinyals}, \citenamefont {Warden}, \citenamefont
  {Wattenberg}, \citenamefont {Wicke}, \citenamefont {Yu},\ and\ \citenamefont
  {Zheng}}]{tensorflow}%
  \BibitemOpen
  \bibfield  {author} {\bibinfo {author} {\bibfnamefont {M.}~\bibnamefont
  {Abadi}}, \bibinfo {author} {\bibfnamefont {A.}~\bibnamefont {Agarwal}},
  \bibinfo {author} {\bibfnamefont {P.}~\bibnamefont {Barham}}, \bibinfo
  {author} {\bibfnamefont {E.}~\bibnamefont {Brevdo}}, \bibinfo {author}
  {\bibfnamefont {Z.}~\bibnamefont {Chen}}, \bibinfo {author} {\bibfnamefont
  {C.}~\bibnamefont {Citro}}, \bibinfo {author} {\bibfnamefont {G.~S.}\
  \bibnamefont {Corrado}}, \bibinfo {author} {\bibfnamefont {A.}~\bibnamefont
  {Davis}}, \bibinfo {author} {\bibfnamefont {J.}~\bibnamefont {Dean}},
  \bibinfo {author} {\bibfnamefont {M.}~\bibnamefont {Devin}}, \bibinfo
  {author} {\bibfnamefont {S.}~\bibnamefont {Ghemawat}}, \bibinfo {author}
  {\bibfnamefont {I.}~\bibnamefont {Goodfellow}}, \bibinfo {author}
  {\bibfnamefont {A.}~\bibnamefont {Harp}}, \bibinfo {author} {\bibfnamefont
  {G.}~\bibnamefont {Irving}}, \bibinfo {author} {\bibfnamefont
  {M.}~\bibnamefont {Isard}}, \bibinfo {author} {\bibfnamefont
  {Y.}~\bibnamefont {Jia}}, \bibinfo {author} {\bibfnamefont {R.}~\bibnamefont
  {Jozefowicz}}, \bibinfo {author} {\bibfnamefont {L.}~\bibnamefont {Kaiser}},
  \bibinfo {author} {\bibfnamefont {M.}~\bibnamefont {Kudlur}}, \bibinfo
  {author} {\bibfnamefont {J.}~\bibnamefont {Levenberg}}, \bibinfo {author}
  {\bibfnamefont {D.}~\bibnamefont {Man\'{e}}}, \bibinfo {author}
  {\bibfnamefont {R.}~\bibnamefont {Monga}}, \bibinfo {author} {\bibfnamefont
  {S.}~\bibnamefont {Moore}}, \bibinfo {author} {\bibfnamefont
  {D.}~\bibnamefont {Murray}}, \bibinfo {author} {\bibfnamefont
  {C.}~\bibnamefont {Olah}}, \bibinfo {author} {\bibfnamefont {M.}~\bibnamefont
  {Schuster}}, \bibinfo {author} {\bibfnamefont {J.}~\bibnamefont {Shlens}},
  \bibinfo {author} {\bibfnamefont {B.}~\bibnamefont {Steiner}}, \bibinfo
  {author} {\bibfnamefont {I.}~\bibnamefont {Sutskever}}, \bibinfo {author}
  {\bibfnamefont {K.}~\bibnamefont {Talwar}}, \bibinfo {author} {\bibfnamefont
  {P.}~\bibnamefont {Tucker}}, \bibinfo {author} {\bibfnamefont
  {V.}~\bibnamefont {Vanhoucke}}, \bibinfo {author} {\bibfnamefont
  {V.}~\bibnamefont {Vasudevan}}, \bibinfo {author} {\bibfnamefont
  {F.}~\bibnamefont {Vi\'{e}gas}}, \bibinfo {author} {\bibfnamefont
  {O.}~\bibnamefont {Vinyals}}, \bibinfo {author} {\bibfnamefont
  {P.}~\bibnamefont {Warden}}, \bibinfo {author} {\bibfnamefont
  {M.}~\bibnamefont {Wattenberg}}, \bibinfo {author} {\bibfnamefont
  {M.}~\bibnamefont {Wicke}}, \bibinfo {author} {\bibfnamefont
  {Y.}~\bibnamefont {Yu}},\ and\ \bibinfo {author} {\bibfnamefont
  {X.}~\bibnamefont {Zheng}},\ }\href {https://www.tensorflow.org/} {\bibinfo
  {title} {{TensorFlow}: Large-scale machine learning on heterogeneous
  systems}} (\bibinfo {year} {2015}),\ \bibinfo {note} {software available from
  tensorflow.org}\BibitemShut {NoStop}%
\bibitem [{\citenamefont {Chollet}\ \emph {et~al.}(2015)\citenamefont {Chollet}
  \emph {et~al.}}]{keras}%
  \BibitemOpen
  \bibfield  {author} {\bibinfo {author} {\bibfnamefont {F.}~\bibnamefont
  {Chollet}} \emph {et~al.},\ }\href@noop {} {\bibinfo {title} {Keras}},\
  \bibinfo {howpublished} {\url{https://keras.io}} (\bibinfo {year}
  {2015})\BibitemShut {NoStop}%
\bibitem [{\citenamefont {Schmidt}\ and\ \citenamefont
  {Lipson}(2009)}]{Schmidt.2009.Science}%
  \BibitemOpen
  \bibfield  {author} {\bibinfo {author} {\bibfnamefont {M.}~\bibnamefont
  {Schmidt}}\ and\ \bibinfo {author} {\bibfnamefont {H.}~\bibnamefont
  {Lipson}},\ }\bibfield  {title} {\bibinfo {title} {{Distilling free-form
  natural laws from experimental data}},\ }\href
  {http://www.sciencemag.org/content/324/5923/81.short} {\bibfield  {journal}
  {\bibinfo  {journal} {Science}\ }\textbf {\bibinfo {volume} {324}},\ \bibinfo
  {pages} {81 } (\bibinfo {year} {2009})}\BibitemShut {NoStop}%
\bibitem [{\citenamefont {Gentile}\ \emph {et~al.}(2021)\citenamefont
  {Gentile}, \citenamefont {Flynn}, \citenamefont {Knauer}, \citenamefont
  {Wiebe}, \citenamefont {Paesani}, \citenamefont {Granade}, \citenamefont
  {Rarity}, \citenamefont {Santagati},\ and\ \citenamefont
  {Laing}}]{gentile2021learning}%
  \BibitemOpen
  \bibfield  {author} {\bibinfo {author} {\bibfnamefont {A.~A.}\ \bibnamefont
  {Gentile}}, \bibinfo {author} {\bibfnamefont {B.}~\bibnamefont {Flynn}},
  \bibinfo {author} {\bibfnamefont {S.}~\bibnamefont {Knauer}}, \bibinfo
  {author} {\bibfnamefont {N.}~\bibnamefont {Wiebe}}, \bibinfo {author}
  {\bibfnamefont {S.}~\bibnamefont {Paesani}}, \bibinfo {author} {\bibfnamefont
  {C.~E.}\ \bibnamefont {Granade}}, \bibinfo {author} {\bibfnamefont {J.~G.}\
  \bibnamefont {Rarity}}, \bibinfo {author} {\bibfnamefont {R.}~\bibnamefont
  {Santagati}},\ and\ \bibinfo {author} {\bibfnamefont {A.}~\bibnamefont
  {Laing}},\ }\bibfield  {title} {\bibinfo {title} {Learning models of quantum
  systems from experiments},\ }\href@noop {} {\bibfield  {journal} {\bibinfo
  {journal} {Nature Physics}\ }\textbf {\bibinfo {volume} {17}},\ \bibinfo
  {pages} {837} (\bibinfo {year} {2021})}\BibitemShut {NoStop}%
\bibitem [{\citenamefont {Losing}\ \emph {et~al.}(2018)\citenamefont {Losing},
  \citenamefont {Hammer},\ and\ \citenamefont
  {Wersing}}]{losing2018incremental}%
  \BibitemOpen
  \bibfield  {author} {\bibinfo {author} {\bibfnamefont {V.}~\bibnamefont
  {Losing}}, \bibinfo {author} {\bibfnamefont {B.}~\bibnamefont {Hammer}},\
  and\ \bibinfo {author} {\bibfnamefont {H.}~\bibnamefont {Wersing}},\
  }\bibfield  {title} {\bibinfo {title} {Incremental on-line learning: A review
  and comparison of state of the art algorithms},\ }\href@noop {} {\bibfield
  {journal} {\bibinfo  {journal} {Neurocomputing}\ }\textbf {\bibinfo {volume}
  {275}},\ \bibinfo {pages} {1261} (\bibinfo {year} {2018})}\BibitemShut
  {NoStop}%
\bibitem [{\citenamefont {Weiss}\ \emph {et~al.}(2016)\citenamefont {Weiss},
  \citenamefont {Khoshgoftaar},\ and\ \citenamefont {Wang}}]{weiss2016survey}%
  \BibitemOpen
  \bibfield  {author} {\bibinfo {author} {\bibfnamefont {K.}~\bibnamefont
  {Weiss}}, \bibinfo {author} {\bibfnamefont {T.~M.}\ \bibnamefont
  {Khoshgoftaar}},\ and\ \bibinfo {author} {\bibfnamefont {D.}~\bibnamefont
  {Wang}},\ }\bibfield  {title} {\bibinfo {title} {A survey of transfer
  learning},\ }\href@noop {} {\bibfield  {journal} {\bibinfo  {journal}
  {Journal of Big data}\ }\textbf {\bibinfo {volume} {3}},\ \bibinfo {pages}
  {1} (\bibinfo {year} {2016})}\BibitemShut {NoStop}%
\bibitem [{\citenamefont {Hoi}\ \emph {et~al.}(2021)\citenamefont {Hoi},
  \citenamefont {Sahoo}, \citenamefont {Lu},\ and\ \citenamefont
  {Zhao}}]{hoi2021online}%
  \BibitemOpen
  \bibfield  {author} {\bibinfo {author} {\bibfnamefont {S.~C.}\ \bibnamefont
  {Hoi}}, \bibinfo {author} {\bibfnamefont {D.}~\bibnamefont {Sahoo}}, \bibinfo
  {author} {\bibfnamefont {J.}~\bibnamefont {Lu}},\ and\ \bibinfo {author}
  {\bibfnamefont {P.}~\bibnamefont {Zhao}},\ }\bibfield  {title} {\bibinfo
  {title} {Online learning: A comprehensive survey},\ }\href@noop {} {\bibfield
   {journal} {\bibinfo  {journal} {Neurocomputing}\ }\textbf {\bibinfo {volume}
  {459}},\ \bibinfo {pages} {249} (\bibinfo {year} {2021})}\BibitemShut
  {NoStop}%
\end{thebibliography}
\end{document}


\title{Supplementary Materials for \\Experimental graybox quantum system identification and control}

\author{Akram Youssry}
\address{Quantum Photonics Laboratory and Centre for Quantum Computation and Communication Technology, RMIT University, Melbourne, VIC 3000, Australia}

\author{Yang Yang}
\address{Quantum Photonics Laboratory and Centre for Quantum Computation and Communication Technology, RMIT University, Melbourne, VIC 3000, Australia}

\author{Robert J. Chapman}
\address{Quantum Photonics Laboratory and Centre for Quantum Computation and Communication Technology, RMIT University, Melbourne, VIC 3000, Australia}
\address{ETH Zurich, Optical Nanomaterial Group, Institute for Quantum Electronics, Department of Physics, 8093 Zurich, Switzerland}

\author{Ben Haylock}
\address{Centre for Quantum Computation and Communication Technology (Australian Research Council), Centre for Quantum Dynamics, Griffith University, Brisbane, QLD 4111, Australia}
\address{Institute for Photonics and Quantum Sciences, SUPA, Heriot-Watt University, Edinburgh EH14 4AS, United Kingdom}

\author{Francesco Lenzini}
\address{Centre for Quantum Computation and Communication Technology (Australian Research Council), Centre for Quantum Dynamics, Griffith University, Brisbane, QLD 4111, Australia}
\address{Institute of Physics, University of Muenster, 48149 Muenster, Germany}

\author{Mirko Lobino}
\address{Centre for Quantum Computation and Communication Technology (Australian Research Council),
Centre for Quantum Dynamics, Griffith University, Brisbane, QLD 4111, Australia}
\address{Department of Industrial Engineering, University of Trento, via Sommarive 9, 38123 Povo, Trento, Italy}
\address{INFN-TIFPA, Via Sommarive 14, I-38123 Povo, Trento, Italy}

\author{Alberto Peruzzo}
\address{Quantum Photonics Laboratory and Centre for Quantum Computation and Communication Technology, RMIT University, Melbourne, VIC 3000, Australia}

\maketitle
\section{Chip fabrication}
Lithium niobate is an ideal choice for a reconfigurable coupled waveguide array due to its large electro-optic coefficient, and consolidated waveguide fabrication technology. The reconfigurable coupled waveguide array devices (Figure S1a) are fabricated in X-Cut bulk lithium niobate (Gooch \& Housego) using the annealed proton exchange technique. The design was optimised for transmission at 808nm using simulations and methods described in \cite{Lenzini:15}. Waveguides were formed with a width of 5.3\ $\mu$m by a proton exchange step through a titanium mask. An exchange depth d$_e$=0.33\ $\mu$m was formed by immersion in pure benzoic acid at 150$^{\circ}$C for 45 minutes. The subsequent annealing was performed in air at 328$^{\circ}$C for 9 hours and 50 minutes. A silicon dioxide layer of 200\ nm is sputtered on top of the lithium niobate before gold electrodes are defined via photolithography and lift-off in a configuration shown in Figure S1b. Electrical connection to the chip is made via wire bonding. 
Waveguide separation at the ends of the device is 127\ $\mu$m to match the pitch of the butt-coupled fibre array. Consequently, a fan-in and fan-out region is required to bring the waveguides into the coupling region where the centre-to-centre waveguide separation is 10\ $\mu$m, chosen to achieve a coupling coefficient of C=140 at 808\ nm. The bends in the fan-in and fan-out regions are defined by cosine curves to maximise bending radius and minimise loss.
\begin{figure}[h]
     \centering
    \includegraphics[scale=0.7]{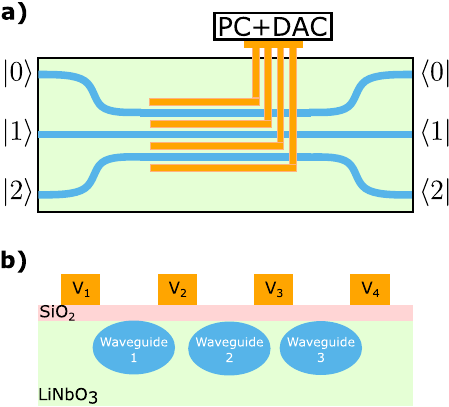}
    \caption{A schematic of the three waveguide chip we utilized in our experiment. a) Top view b) Cross-section}
    \label{chip}
\end{figure}

\section{Chip Modelling}
The Hamiltonian of the photonic chip we use in this paper is typically assumed to be real-valued with this tridiagonal form:
\begin{align}
        H = \begin{pmatrix}
    \beta_1 &  C_{12}   &  0      & \cdots & 0\\
     C_{12} &  \beta_2  & C_{23}  & \cdots & 0\\
     0      & C_{23}    & \beta_3 & \cdots & 0 \\
     \vdots & \vdots    & \vdots  &\ddots  & \vdots\\
     0      & 0         & 0       & \cdots & \beta_{N}
    \end{pmatrix},
    \label{Equ:H_WB}
\end{align}
where each of the diagonal elements  $\beta_i$ represents the propagation constant along the $i^{\text{th}}$ waveguide, while the off-diagonal elements $C_{ij}$ represent the coupling between the neighbouring $i^{\text{th}}$ and $j^{\text{th}}$ waveguides. These Hamiltonian parameters depend on the voltages applied to the electrodes. As first order approximation, the dependence can be considered linear as follows:
\begin{align}
    \beta_i = \beta_i^{(0)} + \Delta\beta_i\Delta V_i,  
    \label{Equ:beta}
\end{align}
where $\Delta V_i$ is the potential difference across the $i^{\text{th}}$ waveguide, and $\beta_i^{(0)}$ and  $\Delta\beta_i$ represent the zero-voltage propagation constant and sensitivity to change in voltage. The coupling coefficients take the form:
\begin{align}
    C_{ij} = C_{ij}^{(0)} + \Delta C_{ij} \left(\Delta V_i +  \Delta V_j\right),
    \label{Equ:C}
\end{align}
where $\Delta V_i$ and $\Delta V_i$ are the potential difference across the $i^{\text{th}}$ and $i^{\text{th}}$ waveguide, and $ C_{ij}^{(0)}$ and $\Delta C_{ij}$ are parameters representing zero-voltage coupling constant and sensitivity to change in voltage. Therefore, there is a total of $4N-2$ trainable parameters to completely specify this Hamiltonian. The length of the waveguides $l$ is fixed and known from the design of the chip, and is equivalent to time for the evolution. 
This model is based on what is widely known and accepted by the photonics community. In particular, the assumption of a real-valued Hamiltonian is justified by the fact that the propagation constant and the coupling coefficient are real for lossless devices \cite{salehText}. The tri-diagonal form is justified by the fact that only nearest-neighbor coupling between waveguides is significant, and the coupling decays exponentially for further waveguides \cite{Bromberg_2009,Peruzzo.2010.Science.10.1126/science.1193515j8}. Finally, the linear dependence on voltage is due to the Pockels effect where the change in the refractive index of the material is proportional to the amplitude of the applied electric field, as shown in \cite{salehText}.

In addition, there exists an effective fan-in and fan-out sections of the chip where the waveguides are gradually separated to the input and output ports of the chip. This effectively induces extra coupling, and the complete device can be modeled as a cascade of three unitaries: the fan-in unitary, the voltage-controlled unitary, and the fan-out unitary. In other words,  
\begin{align}
    U = U_{\text{fan-out}} U(V) U_{\text{fan-in}},
    \label{equ:U}
\end{align}
where $U(V)$ is the voltage-dependent evolution unitary, and each of the fan-in and fan-out unitaries take the form of $e^{-i H_{\text{fan}} l_{\text{fan}}}$. The fan Hamiltonian takes the same tri-diagonal form of the chip Hamiltonian, but with one main difference that it is voltage independent. The fan length $l_{\text{fan}}$ is chosen arbitrarily to be $1$ (since it can be absorbed in fan Hamiltonian). The parameters of the fan Hamiltonian alongside the chip Hamiltonian parameters form the set parameters of the physical model of the chip.

We choose the initial states to be the computational basis states $\{\ket{0}, \ket{1}, \cdots \ket{N}$, where the $\ket{j}$ state encodes a single photon entering the chip at port $j$. Equivalently, if classical laser is used, then the state corresponds to light entering through input port $j$. Therefore, we actually measure $N$ power distributions corresponding to each input basis states. In other words, we measure the probability $P_{i \to j}$ of transitioning from input port $i$ to output port $j$. This gives a total of $N^2$ outputs. For the purposes of this paper, these measurements are sufficient. However, in the absence of any phase measurements, the model predictions are be accurate only in terms of probability amplitudes of the output state (or equivalently the square amplitudes of the unitary describing the chip). One way to measure phase shifts is to use a Mach-Zehnder interferometer to convert phases into powers. In this case, there will be $2N^2$ outputs.

\section{Protocol Implementation}
We collected a dataset of 7000 examples for training, and 1000 for testing. Each example consists of the voltages applied to four electrodes, chosen randomly in the interval $[-1,1]$, and the corresponding measured output power distribution. The electrodes are activated simultaneously in all experiments. Thus, the models we use will have four inputs, and 9 outputs. 

Because the evolution of the system is unitary, and the input states are orthogonal, the output states must also be orthogonal. Particularly the matrix $|U|^2$, which is the unitary after taking squared absolute value element-wise must form a bistochastic matrix. In other words, the columns must be normalized to $1$, and the rows as well. Both the matrix and its inverse  $U^{-1} = U^{\dagger}$ (backward evolution in time) are unitary, and thus the extra requirement of row normalization. While measuring the probability amplitudes of an evolved state, the column normalization is automatically done so that it yields a valid probability distribution. However, the row normalization is not naturally guaranteed due to the presence of experimental noise. Therefore, the measured power distributions are further post-processed to ensure the normalization holds for both rows and columns. We use iterated proportional fitting algorithm which is a standard procedure in statistics. First we arrange the measured power distributions into an $3\times 3$ matrix where each column corresponds to the output of a particular input state. The idea simply is to keep alternating between normalizing rows and normalizing columns until convergence, which is theoretically guaranteed \cite{sinkhorn1967concerning}. Finally, the columns are stacked together to form an $9 \times 1$ vector that matches the model output. The procedure for normalizing the power distributions is done for each example of the dataset. It is also done for the control dataset. 

For the GB model, we implemented the structure described in the main text of the paper. In particular, the neural network consists of three layers of 50, 100, and 18 neurons respectively. The first two layers have a hyperbolic tangent activation, while the last one has linear activation. The whitebox parts consists of standard quantum evolution and measurement operations. Because we do not impose any constraints of the Hamiltonian structure, the interpretation is that it is an effective Hamiltonian that results in the overall unitary (i.e. chip unitary and fan unitaries) as described in Equation \ref{equ:U}. In fact any unitary can be written as the imaginary evolution of some Hermitian operator. 

For the fully WB model, we implement the physical model described by Equations \ref{Equ:H_WB},\ref{Equ:beta}, \ref{Equ:C}, \ref{equ:U} setting $N=3$. Thus, the first layer of the ML model implements the set of parameterized equations \ref{Equ:beta}, \ref{Equ:C}. The second layer in the model, is the Hamiltonian construction where the Hamiltonian parameters (that are the output of the first layer), are arranged into an $3\times 3$ matrix as described by Equation \ref{Equ:H_WB}. This layer is followed by the quantum evolution layer to calculate the voltage dependent unitary. After that, we construct a layer that implements the overall evolution by including the fan-in and fan-out of the waveguides, implementing Equation \ref{equ:U}. 

Finally, for the fully BB model, we have three layers. The first layer consists of 50 neurons with hyperbolic tangent activation function. The second layer consists of 100 neurons of hyperbolic tangent activation as well. The final output layer consists of three concatenated layers of 3 neurons with softmax activation.

We then trained the three models using the Adam optimizer \cite{Adam} (which is one of the most commonly-used and successful methods for training NN) with learning rate 0.003 for 3000 iterations. 
For each of the three ML models, we run the output controller as well as the unitary controller for 1000 random states and unitaries. 
After obtaining the optimal controls for all cases, we apply them experimentally on the chip, and measure the corresponding output distributions for all initial states. We also the same post-processing step that is used for the training and testing datasets. Since we restrict the measurements in this paper to computational basis, we can use the classical fidelity $F(p,q) = \sum_i \sqrt{p_i}\sqrt{q_i}$ between probability distributions as a metric of how close the measured distributions are to the target ones. Since we have 3 initial states, we get three fidelities for each target example, so we take the average over the three initial states. We repeat this for each of the 1000 targets, and perform statistical analysis on the calculated fidelities. For each of the three models considered, we report below the statics distributions for each separate input. It can be note that the performance slightly change between the inputs.

\begin{figure*}[h]
    \centering
    \begin{subfigure}[b]{\textwidth}
        \includegraphics[scale=0.44]{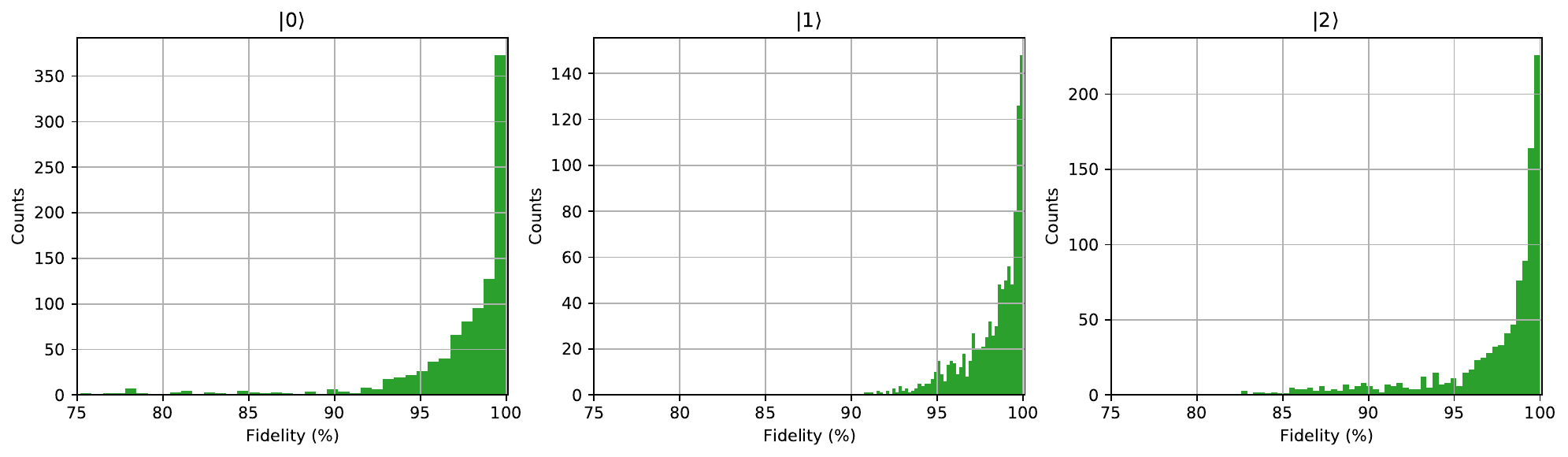}
        \caption{Whitebox} 
    \end{subfigure}\\
    
    \begin{subfigure}[b]{\textwidth}
        \includegraphics[scale=0.44]{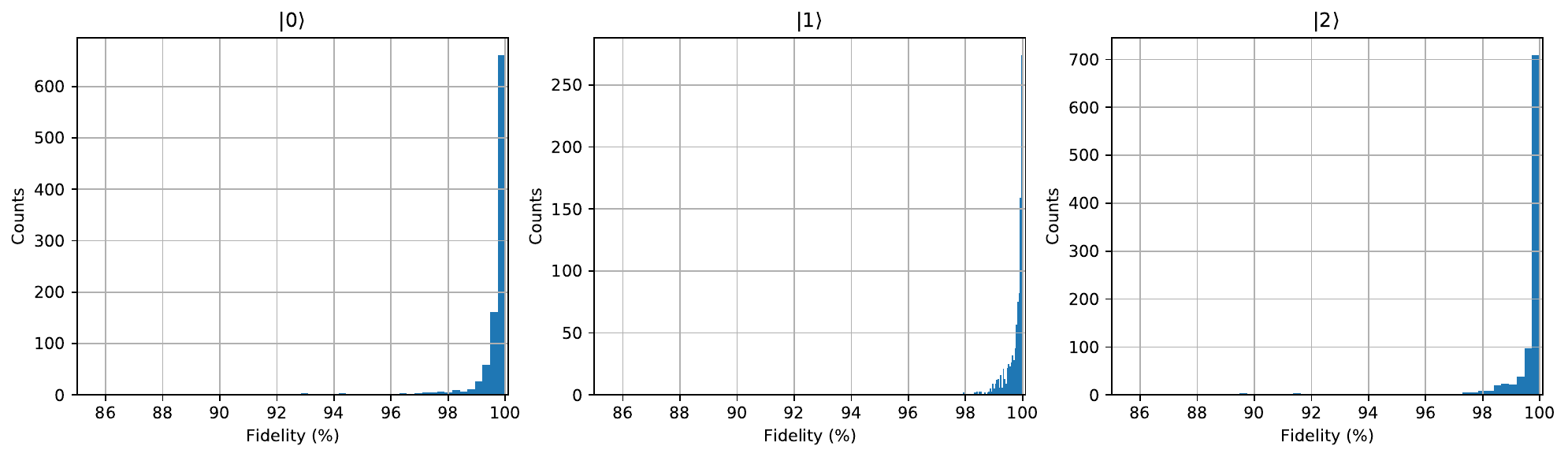}
        \caption{Blackbox} 
    \end{subfigure}\\
    
    \begin{subfigure}[b]{\textwidth}
        \includegraphics[scale=0.44]{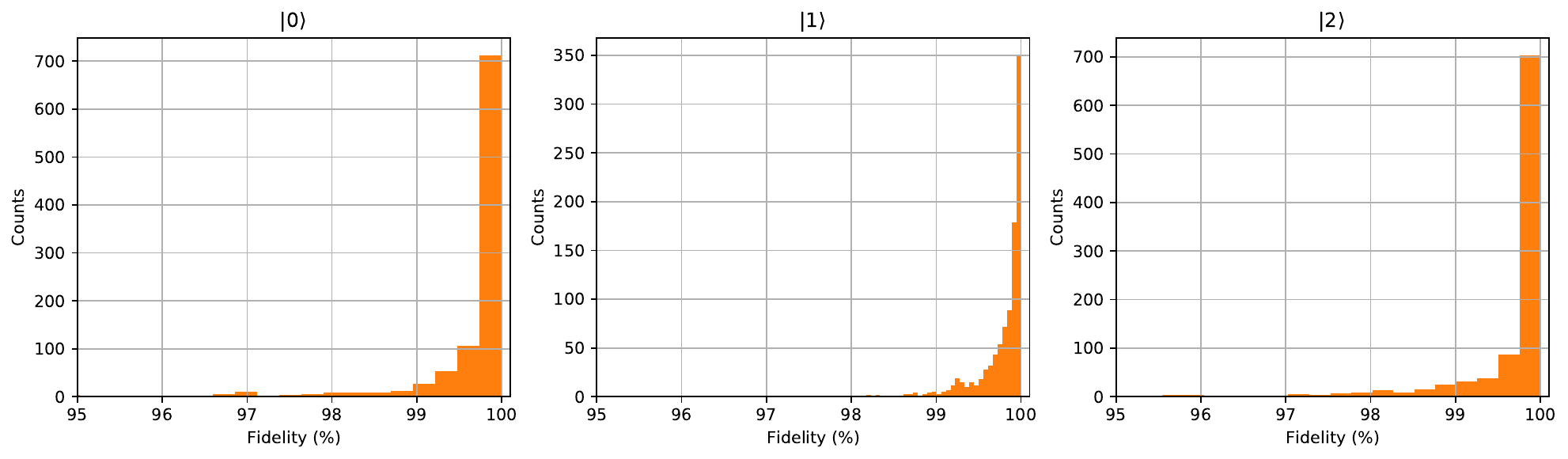}
        \caption{Graybox} 
    \end{subfigure}
    \caption{The distribution of fidelity between the experimentally measured output power distribution and the desired target distribution. The rows show the results for each of the whitebox (WB), blackbox (BB), and graybox (GB) respectively. The WB model is based on a tri-diagonal real-valued Hamiltonian with linear dependence on voltages.} The three columns represent different input states.
    \label{fig:fidelity}
\end{figure*}

\section{Simulations}
We implemented the proposed protocol on a synthetic simulated dataset to study the performance of the different ML models in an ideal scenario. For the simulator, we considered a Hamiltonian with an extra quadratic term in voltage to model the possibility of having a non-linearity that is not known for a WB physical model. Additionally, we considered measuring phase information and recording it as a part of the dataset. This is done by doing interferometer measurements on the outputs of the chip as described in \cite{youssry2020modeling}. It is possible to infer the magnitude and phases of the output state if we did two interferometer measurements with different angles for each input/output pair. This will result in double the number of outputs (i.e. $2N^2$) which is the requirement for completely determining all the elements of the evolution unitary. We matched all the settings (such as dataset size, and ML hyperparameters) as the those used for the experimental data in the paper. The BB model however was modified in the output layer where a sigmoid activation was used instead of softmax, as we no longer require the normalization of the outputs since they now represent the interference measurements.

\begin{figure*}
    \centering
    \includegraphics[scale=0.48]{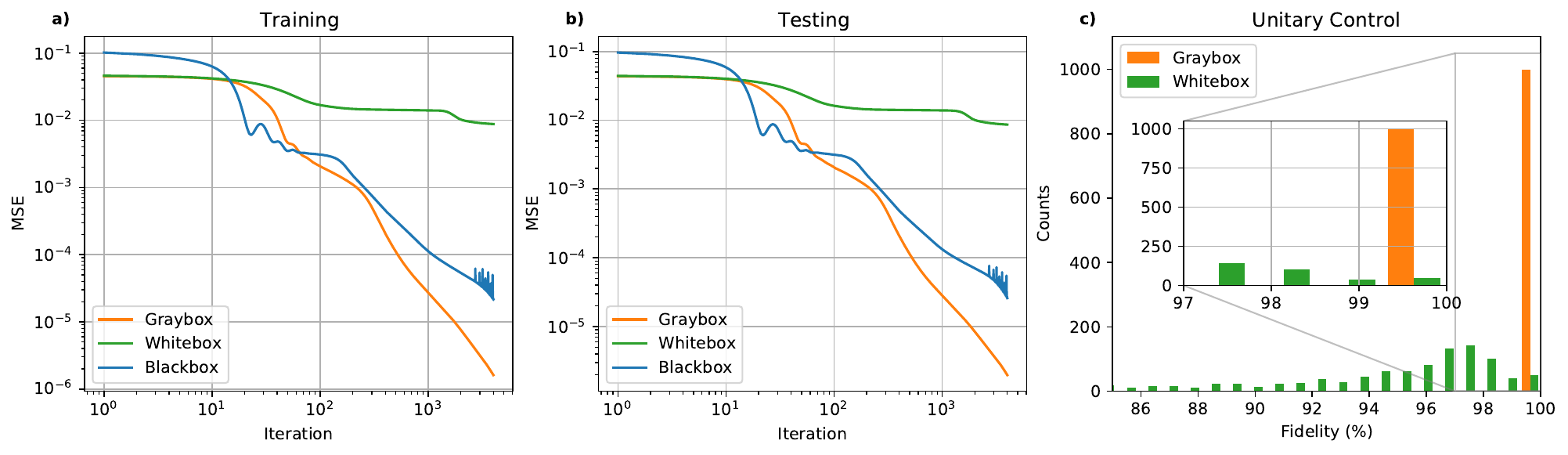}
    
    \caption{Results of training the different model on the simulated dataset for the 3-mode chip. The WB model is based on a tri-diagonal real-valued Hamiltonian with linear dependence on voltages. The MSE is plotted versus iteration number for a) training set, and b) for testing set. c) The distribution of gate fidelity between the simulated unitary and the desired target unitary for each of the whitebox and graybox controllers.}
    \label{fig:sim}
\end{figure*}

Figures \ref{fig:sim}a, \ref{fig:sim}b show the results of the training and testing the three ML architectures against the simulated dataset. It clearly shows that the WB had the worst performance because it is based on a physical model where the Hamiltonian depends linearly on the voltage whereas the simulated chip has a quadratic term as well. The interesting observation though is that at the end of iterations, the GB performed much better than the BB for the same number of iterations. This motivates the idea that with the GB structure, it is easier for the machine to learn the dataset because we already provide part of the dynamical equations. Fig. \ref{fig:sim}c shows the results for comparing the gate fidelity using the WB and GB. As expected the GB outperforms the WB because it models the data more accurately and so it is expected to have better control. In conclusion, the simulation results agree with the experimental results, and also shows the applicability of the proposed method for a different and more complex setup.

\begin{figure*}
    \centering
    \includegraphics[scale=0.48]{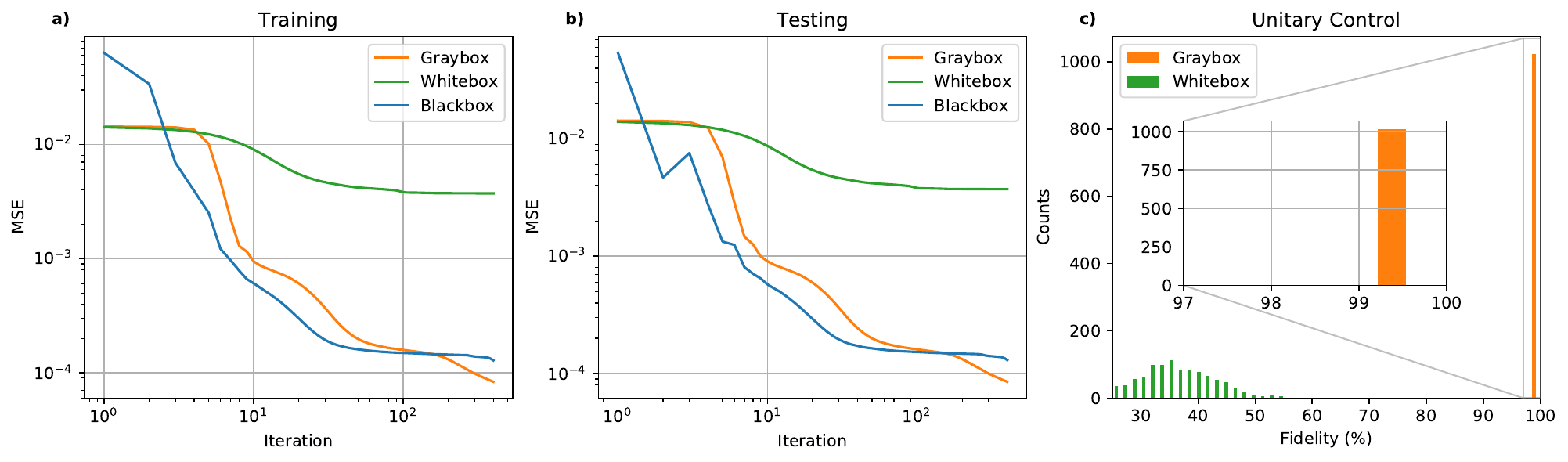}
    \caption{Results of training the different models on the simulated dataset for a 32-mode chip. The WB model is based on a tri-diagonal real-valued Hamiltonian with linear dependence on voltages. The MSE is plotted versus iteration number for a) training set, and b) testing set. c) The distribution of gate fidelity between the simulated unitary and the desired target unitary for each of the whitebox and graybox controllers.}
    \label{fig:sim32}
\end{figure*}

Additionally, we performed another simulation for a 32-mode chip, to study the applicability of the proposed method to much higher-dimensional system. The dataset consisted of 65536 examples for training, and 8192 for testing. The layers of the neural networks had 100 and 300 nodes for the hidden layers.  For assessing control, we utilized 1024 examples out of the testing set. Figure \ref{fig:sim32} shows the training, testing, and unitary control performance. The GB maintains its performance compared to BB and WB. Comparing to the $3\times3$ case, the training and testing MSE are higher. However, the unitary control in the GB case still shows most examples are concentrated around 99\% of fidelity.

\section{Re-designing the Whitebox}
The results of our experiments, particularly from the GB,  show that the best way to model the chip is using a general $ N\times N$ complex-valued Hermitian Hamiltonian that is non-linearly dependent on the control voltages. This deviates significantly from the standard WB model that uses the tri-diagonal real-valued linearly voltage-dependent Hamiltonian. In this section, we explore the effect of relaxing the WB assumptions based on the new knowledge acquired from the GB results. 

\begin{figure*}
    \centering
    \includegraphics[scale=0.48]{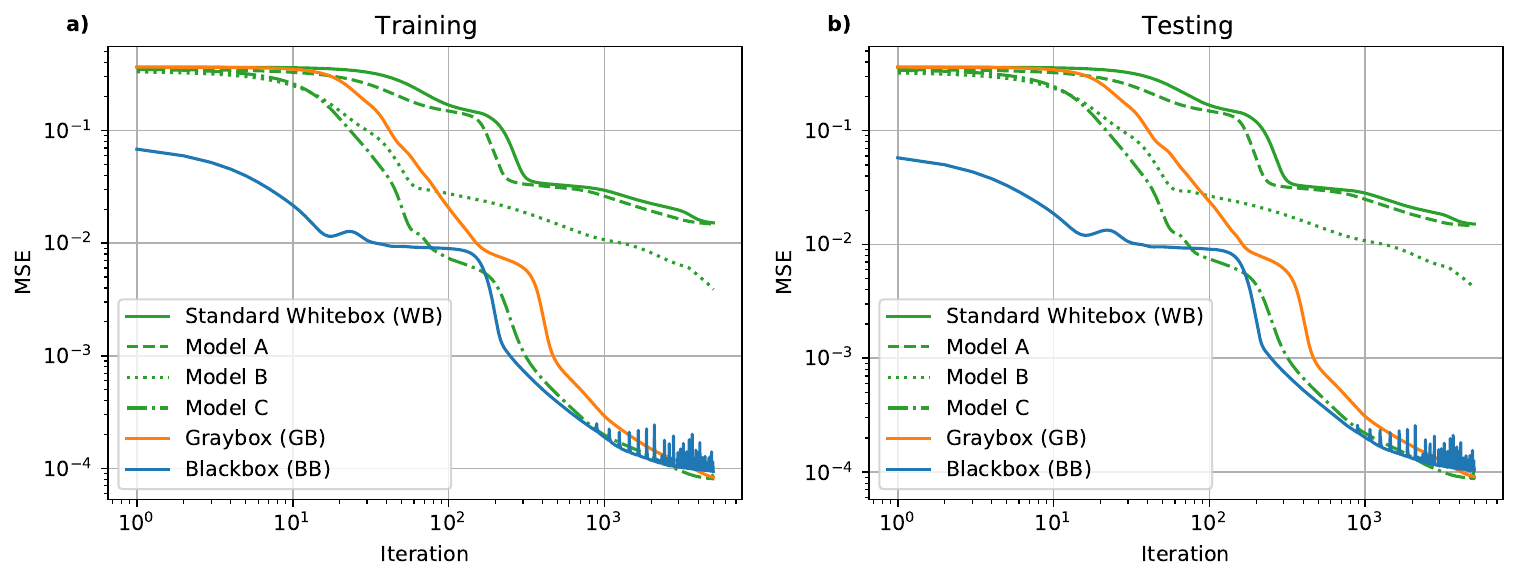}
    \caption{Results of training the different models on the experimental dataset. The MSE is plotted versus iteration number for a) training set, and b) testing set. Here the WB model is the original real-tridiagonal and linear. Model A is the modification with a complex-valued Hamiltonian, Model B is the modification with complex-valued and non-tridiagonal Hamiltonian, and finally Model C is with complex-valued non-tridiagonal and non-linear Hamiltonian. The 4 WB models separate the fan-in/fan-out from the reconfigurable section. The GB model does not separate the sections, and also uses a complex non-tridiagonal and non-linear Hamiltonian.}
    \label{fig:WB_Models}
\end{figure*}

Here we introduce three additional WB models:
\begin{enumerate}[A)]
    \item complex, tri-diagonal, and linear voltage dependence,
    \item complex, non-tridiagonal, and linear voltage dependence,
    \item complex, non-tridigonal, and non-linear voltage dependence (using an NN)
\end{enumerate}
For all models (including the original WB in the main text), we separate the fan-in and fan-out Hamiltonians (that are voltage independent), so that we can study the reconfigurable part separately. We use the same mathematical structure for the fan-in and fan-out Hamiltonians as the reconfigurable part. Next, we fit those models with the experimental data, and compare against the standard WB and the GB. Figure \ref{fig:WB_Models} shows the MSE for the training and testing datasets. The results show that relaxing the the assumptions on the WB, the MSE improves. Model C (with least assumptions) performs comparably to the GB towards the end of the iterations. 

The fact that the complex non-tridiagonal fits better implies that we are fitting an effective Hamiltonian of non-commuting sections, or more generally a time-dependent Hamiltonian. In other words, the model is trying to fit
\begin{align}
     H_{\text{eff}}(\mathbf{V}) = \frac{1}{-i L}\log{\left(\mathcal{T}_{+} e^{-i\int_0^L{ H(\mathbf{V}, z)dz}} \right)},
    \label{equ:H_eff}
\end{align}
where $\mathcal{T}_{+}$ is the time-ordering operator, and $H(\mathbf{V}, z)$ is the physical Hamiltonian that depends on the control voltage $\mathbf{V}$ as well as position $z$ along the propagation direction. Since we already separated the fan-in and fan-out from the reconfigurable part, we can conclude that the reconfigurable part itself is varying along the propagation direction. For example, the electrodes might not be symmetric due to fabrication imperfections.

As for the non-linear dependence on the voltage, this can be attributed to multiple sources. First, it can be physical due to higher-order Pockel's effect, although it is unlikely in our experiment as we limit the maximum control voltage. Alternatively, the non-linearity could be a mathematical artifact of the optimization process. By looking back at Equation \ref{equ:H_eff}, we see that the effective Hamiltonian will be likely non-linear because of the matrix logarithm and time-ordered evolution operations. Moreover, due to the non-uniqueness of Hamiltonians generating a given unitary, we can have a case where both a linear and a non-linear Hamiltonian would generate exactly the same unitary. We illustrate this with a $2\times 2$ system as an example.
Let $H_1(\mathbf{V})$ be a $2\times 2$ Hamiltonian of the form
\begin{align*}
     H_1(\mathbf{V}) &= \begin{pmatrix}f_1(\mathbf{V}) & g_1(\mathbf{V}) \\ g_1(\mathbf{V}) & -f_1(\mathbf{V}) \end{pmatrix}
\end{align*}
where $f_1(\mathbf{V})$ and $g_1(\mathbf{V})$ are two linear functions of the control voltage $\mathbf{V}$. We want to find another equivalent Hamiltonian $H_2$ of the same form
\begin{align*}
     H_2(\mathbf{V}) &= \begin{pmatrix}f_2(\mathbf{V}) & g_2(\mathbf{V}) \\ g_2(\mathbf{V}) & -f_2(\mathbf{V}) \end{pmatrix}
\end{align*}
i.e. they generate the same unitaries $U_1$ and $U_2$, such that, $f_2(\mathbf{V})$ and $g_2(\mathbf{V})$ are now non-linear functions of $\mathbf{V}$. For this form of the Hamiltonian, we can express the evolved unitary easily if we rewrite the Hamiltonian in the Pauli vector notation
\begin{align*}
    A &= a (\hat{n} \cdot \vec{\sigma}) \\
    U &= e^{i A} = I \cos(a) + i \sin(a) (\hat{n} \cdot \vec{\sigma}),
\end{align*}
where $a$ is a scalar, $I$ is the identity matrix, $\hat{n}:=[n_x, n_y, n_z]^T$ is a three-dimensional unit vector, and $\hat{n} \cdot \vec{\sigma}:= n_x \sigma_x + n_y \sigma_y + n_z \sigma_z$, $\sigma_j$ is the Pauli matrix along the $j^{\text{th}}$ direction. Thus, rewriting the two Hamiltonians in that form we have,
\begin{widetext}
\begin{align*}
    H_1(\mathbf{V}) &= \sqrt{f_1(\mathbf{V})^2 + g_1(\mathbf{V})^2}\left(\frac{g_1(\mathbf{V})}{\sqrt{f_1(\mathbf{V})^2 + g_1(\mathbf{V})^2}}\sigma_x + \frac{f_1(\mathbf{V})}{\sqrt{f_1(v)^2 + g_1(\mathbf{V})^2}}\sigma_z\right)\\
    H_2(\mathbf{V})&= \sqrt{f_2(\mathbf{V})^2 + g_2(\mathbf{V})^2}\left(\frac{g_2(\mathbf{V})}{\sqrt{f_2(\mathbf{V})^2 + g_2(\mathbf{V})^2}}\sigma_x + \frac{f_2(\mathbf{V})}{\sqrt{f_2(\mathbf{V})^2 + g_2(\mathbf{V})^2}}\sigma_z\right).
\end{align*}
The evolution can be calculated then as 
\begin{align*}
    U_1&= I \cos{\left(\sqrt{f_1(\mathbf{V})^2 + g_1(\mathbf{V})^2}L\right)} - i \sin{\left(\sqrt{f_1(\mathbf{V})^2 + g_1(\mathbf{V})^2}L\right)}\left(\frac{g_1(\mathbf{V})}{\sqrt{f_1(\mathbf{V})^2 + g_1(\mathbf{V})^2}}\sigma_x + \frac{f_1(\mathbf{V})}{\sqrt{f_1(v)^2 + g_1(\mathbf{V})^2}}\sigma_z\right)\\
    U_2&= I \cos{\left(\sqrt{f_2(\mathbf{V})^2 + g_2(\mathbf{V})^2}L\right)} - i \sin{\left(\sqrt{f_2(\mathbf{V})^2 + g_2(\mathbf{V})^2}L\right)}\left(\frac{g_2(\mathbf{V})}{\sqrt{f_2(\mathbf{V})^2 + g_2(\mathbf{V})^2}}\sigma_x + \frac{f_2(\mathbf{V})}{\sqrt{f_2(\mathbf{V})^2 + g_2(\mathbf{V})^2}}\sigma_z\right)\\
\end{align*}
Equating the two unitaries, we obtain the three conditions to satisfy:
\begin{align*}
     \cos{\left(\sqrt{f_1(\mathbf{V})^2 + g_1(\mathbf{V})^2}L\right)} =  \cos{\left(\sqrt{f_2(\mathbf{V})^2 + g_2(\mathbf{V})^2}L\right)}  &\implies 
     \sqrt{f_1(\mathbf{V})^2 + g_1(\mathbf{V})^2} = \sqrt{f_2(\mathbf{V})^2 + g_2(\mathbf{V})^2} + \frac{2\pi k}{L}, \quad k\in \mathbb{Z} \\
      \frac{f_1(\mathbf{V})}{\sqrt{f_1(\mathbf{V})^2 + g_1(\mathbf{V})^2}} = \frac{f_2(v)}{\sqrt{f_2(\mathbf{V})^2 + g_2(\mathbf{V})^2}} &\implies 
      \boxed{f_2(\mathbf{V}) = f_1(\mathbf{V})\left(1 - \frac{2\pi k}{\sqrt{f_1(\mathbf{V})^2 + g_1(\mathbf{V})^2}L}\right), \quad k\in \mathbb{Z}}\\
      \frac{g_1(\mathbf{V})}{\sqrt{f_1(\mathbf{V})^2 + g_1(\mathbf{V})^2}} = \frac{g_2(v)}{\sqrt{f_2(\mathbf{V})^2 + g_2(\mathbf{V})^2}} &\implies \boxed{g_2(\mathbf{V}) = g_1(\mathbf{V})\left(1 - \frac{2\pi k}{\sqrt{f_1(\mathbf{V})^2 + g_1(\mathbf{V})^2}L}\right),\quad k\in \mathbb{Z}}.
\end{align*}
\end{widetext}
We can see that by exploiting the periodicity of the cosine, we can obtain a non-linear solution for $f_2(\mathbf{V})$ and $f_2(\mathbf{V})$. Thus, there exist an infinite family of transformations parameterized by an integer $k$, that define an equivalent class of Hamiltonians in terms of the evolved unitary. The general concept can be extend to larger systems, while it can verified numerically, finding an analytical expression would be very difficult.

The consequence of this model invariance, and the fact that we cannot directly measure the Hamiltonian, but only the unitary encoded as power distributions, means that there is no way to decide from experimental data whether a linear or a non-linear model is the groundtruth. Moreover, the NN is more likely to find a non-linear solution simply because there are infinitely many of them, and there is no preference for one over the other. Therefore, given the experimental data and with the use of ML optimization techniques, it is really difficult to assert the source of non-linearity. Nevertheless, this does not affect anything related to the purpose of system identification and control that we focus on in this paper.

\color{black}